\documentclass[preprint,a4paper]{elsarticle}

\usepackage{version}
\usepackage{amsmath}
\usepackage{amssymb}
\usepackage{fullpage}
\usepackage{booktabs}
\usepackage{longtable}
\usepackage{tikz}
\usepackage{hyperref}
\usepackage[capitalise]{cleveref}
\usepackage[vlined]{algorithm2e}
\usepackage{subcaption}

\usepackage{listings}

\hypersetup{
  colorlinks
}

\crefname{algocf}{Algorithm}{Algorithms}
\Crefname{algocf}{Algorithm}{Algorithms}
\crefname{lstlisting}{Listing}{Listings}
\Crefname{lstlisting}{Listing}{Listings}

\crefname{condition}{Condition}{Conditions}
\Crefname{condition}{Condition}{Conditions}

\crefformat{appendix}{#2#1#3}

\newcommand{\propA}{Sequential Bound}
\newcommand{\propB}{Non-increasing Runtimes}
\newcommand{\propC}{Repeatability}

\lstnewenvironment{haskellcode}[2]
{
    \lstset{
      language=Haskell,
      numbers=left,
      float,
      caption=#1,
      captionpos=b,
      label=#2,
      basicstyle=\small\ttfamily,
      flexiblecolumns=false,
      basewidth={0.5em,0.45em},
      moredelim=**[is][\color{red}]{@}{@},
      literate={+}{{$+$}}1 {/}{{$/$}}1 {*}{{$*$}}1 {=}{{$=$}}1
               {>}{{$>$}}1 {<}{{$<$}}1 {\\}{{$\lambda$}}1
               {\\\\}{{\char`\\\char`\\}}1
               {->}{{$\rightarrow$}}2 {>=}{{$\geq$}}2 {<-}{{$\leftarrow$}}2
               {<=}{{$\leq$}}2 {=>}{{$\Rightarrow$}}2
               {\ .}{{$\circ$}}2 {\ .\ }{{$\circ$}}2
               {>>}{{>>}}2 {>>=}{{>>=}}2
               {|}{{$\mid$}}1
    }}
  {}
\lstnewenvironment{haskellcodenofloat}
{
    \lstset{
      language=Haskell,
      captionpos=b,
      basicstyle=\small\ttfamily,
      flexiblecolumns=false,
      basewidth={0.5em,0.45em},
      moredelim=**[is][\color{red}]{@}{@},
      literate={+}{{$+$}}1 {/}{{$/$}}1 {*}{{$*$}}1 {=}{{$=$}}1
               {>}{{$>$}}1 {<}{{$<$}}1 {\\}{{$\lambda$}}1
               {\\\\}{{\char`\\\char`\\}}1
               {->}{{$\rightarrow$}}2 {>=}{{$\geq$}}2 {<-}{{$\leftarrow$}}2
               {<=}{{$\leq$}}2 {=>}{{$\Rightarrow$}}2
               {\ .}{{$\circ$}}2 {\ .\ }{{$\circ$}}2
               {>>}{{>>}}2 {>>=}{{>>=}}2
               {|}{{$\mid$}}1
    }}
  {}

\usetikzlibrary{shapes.multipart}
\usetikzlibrary{shapes.geometric}

\definecolor{safelightblue}{rgb}{0.65098, 0.807843, 0.890196}
\definecolor{safedarkblue}{rgb}{0.121569, 0.470588, 0.705882}
\definecolor{safeverylightblue}{rgb}{0.857843, 0.931961, 0.996078}
\definecolor{safeverydarkblue}{rgb}{0.007843, 0.219608, 0.345098}
\definecolor{safelightorange}{rgb}{0.996078, 0.931961, 0.857843}
\definecolor{safelightgreen}{rgb}{0.698039, 0.87451, 0.541176}
\definecolor{safemediumorange}{rgb}{0.74902, 0.505882, 0.490196}
\definecolor{safelightgreen}{rgb}{0.698039, 0.87451, 0.541176}
\definecolor{safedarkgreen}{rgb}{0.2, 0.627451, 0.172549}
\definecolor{safepurple}{rgb}{0.458824, 0.439216, 0.701961}
\definecolor{safedarkorange}{rgb}{0.34902, 0.176471, 0.015686}
\definecolor{safelightgrey}{rgb}{0.7, 0.7, 0.7}
\definecolor{safenearlywhite}{rgb}{0.9, 0.9, 0.9}
\definecolor{safereallynearlywhite}{rgb}{0.93, 0.93, 0.93}

\tikzset{vertex/.style={draw, circle, inner sep=0pt, minimum size=0.5cm, font=\small\bfseries}}
\tikzset{notvertex/.style={vertex, color=white, text=black}}
\tikzset{plainvertex/.style={vertex}}
\tikzset{selectedvertex/.style={vertex, fill=safedarkblue}}
\tikzset{delvertex/.style={vertex, dotted, color=safelightgrey}}
\tikzset{vertexc1/.style={vertex, fill=safelightblue}}
\tikzset{vertexc2/.style={vertex, fill=safedarkblue, text=safenearlywhite}}
\tikzset{vertexc3/.style={vertex, fill=safelightorange}}
\tikzset{vertexc4/.style={vertex, fill=safedarkorange, text=safenearlywhite}}
\tikzset{vertexc5/.style={vertex, fill=safedarkgreen}}

\tikzset{edge/.style={color=safelightgrey}}
\tikzset{bedge/.style={ultra thick}}
\tikzset{deledge/.style={dotted, color=safelightgrey}}
\tikzset{edgel1/.style={color=safeverydarkblue}}
\tikzset{edgel2/.style={color=safemediumorange}}
\tikzset{edgel3/.style={ultra thick, color=safedarkblue}}
\tikzset{edgel4/.style={ultra thick, color=safelightorange}}

\tikzset{processarrow/.style={->, very thick, decorate, decoration={snake, post length=0.5mm}}}
\tikzset{brace/.style={decorate, decoration={brace}, thick}}
\tikzset{bracem/.style={decorate, decoration={brace, mirror}, thick}}
\tikzset{label/.style={font=\small}}

\tikzset{choice/.style={rectangle split,
               rectangle split horizontal,
               rectangle split parts=#1,
               draw,
               anchor=center}}

\tikzset{moreTree/.style={isosceles triangle,
               draw,
               anchor=north,
               shape border rotate=90}}

\begin{document}

\title{Replicable Parallel Branch and Bound Search\tnoteref{t1}}
\tnotetext[t1]{\textcopyright\ 2017. This manuscript version is made available under the CC-BY-NC-ND 4.0 license \url{http://creativecommons.org/licenses/by-nc-nd/4.0/}}

\author[UoG]{Blair Archibald\corref{cor}}
\ead{b.archibald.1@research.gla.ac.uk}
\author[UoG]{Patrick Maier}
\ead{Patrick.Maier@glasgow.ac.uk}
\author[UoG]{Ciaran McCreesh}
\ead{c.mccreesh.1@research.gla.ac.uk}
\author[HW]{Robert Stewart}
\ead{r.stewart@hw.ac.uk}
\author[UoG]{Phil Trinder}
\ead{Phil.Trinder@glasgow.ac.uk}

\address[UoG]{School Of Computing Science, University of Glasgow, Scotland, G12 8QQ}
\address[HW]{Heriot-Watt University, Edinburgh, Scotland, EH14 4AS}
\cortext[cor]{Corresponding author}

\begin{abstract}
  Combinatorial branch and bound searches are a common technique for
  solving global optimisation and decision problems. Their performance
  often depends on good search order heuristics, refined over decades
  of algorithms research.  Parallel search necessarily deviates from
  the sequential search order, sometimes dramatically and
  unpredictably, e.g. by distributing work at random.  This can
  disrupt effective search order heuristics and lead to unexpected and
  highly variable parallel performance.  The variability makes it hard
  to reason about the parallel performance of combinatorial searches.

  This paper presents a generic parallel branch and bound skeleton,
  implemented in Haskell, with replicable parallel performance.  The
  skeleton aims to preserve the search order heuristic by distributing
  work in an ordered fashion, closely following the sequential search
  order.  We demonstrate the generality of the approach by applying
  the skeleton to 40 instances of three combinatorial problems:
  Maximum Clique, 0/1 Knapsack and Travelling Salesperson.  The
  overheads of our Haskell skeleton are reasonable: giving slowdown
  factors of between 1.9 and 6.2 compared with a class-leading,
  dedicated, and highly optimised C++ Maximum Clique solver.  We
  demonstrate scaling up to 200 cores of a Beowulf cluster, achieving
  speedups of 100x for several Maximum Clique instances.  We
  demonstrate low variance of parallel performance across all
  instances of the three combinatorial problems and at all scales up
  to 200 cores, with median Relative Standard Deviation (RSD) below
  2\%.  Parallel solvers that do not follow the sequential search
  order exhibit far higher variance, with median RSD exceeding 85\%
  for Knapsack.
\end{abstract}


\maketitle

\section{Introduction}
\label{sec:introduction}

Branch and bound backtracking searches are a widely used class of algorithms.
They are often applied to solve a range of NP-hard optimisation problems
such as integer and non-linear programming problems;
important applications include frequency planning in cellular networks
and resource scheduling, e.g. assigning deliveries
to routes~\cite{lamporte-VRPSurvey}.

Branch and bound systematically explores a \emph{search tree} by sub-dividing
the search space and \emph{branching} recursively into each sub-space.
The advantage of branch and bound over exhaustive enumeration stems from the
way branch and bound \emph{prunes} branches that cannot better the
\emph{incumbent}, i.e.\ the current best solution, potentially drastically
reducing the number of branches to be explored.

The effectiveness of pruning depends on two factors:
1) the accuracy of the problem-specific heuristic to compute \emph{bounds}
2) the value of optimal solutions in each branch, and on the quality of the
incumbent; the closer to optimal the incumbent, the more can be pruned.
As a result, branch and bound is sensitive to \emph{search order},
i.e. to the order in which branches are explored.

A good search order can improve the performance of branch and bound
dramatically by finding a good incumbent early on, and highly optimised
sequential algorithms following the branch and bound paradigm often
rely on very specific orders for performance.

Branch and bound algorithms are hard to parallelise for a number of reasons.
Firstly, while branching creates opportunities for speculative parallelism where
multiple \emph{workers} i.e\ threads/processors search particular branches in
parallel, pruning counteracts this, limiting potential parallelism. Secondly,
parallel pruning requires processors sharing access to the incumbent, which
limits scalability. Thirdly, parallel exploration of irregularly shaped search
trees generates unpredictable numbers of parallel tasks, of highly variable
duration, posing challenges for task scheduling. Finally, and most importantly,
parallel exploration alters the search order, potentially impacting the
effectiveness of pruning.

As a result of the last point in particular, parallel branch and bound searches
can exhibit unusual performance characteristics. For instance, slowdowns can
arise when the sequential search finds an optimal incumbent quickly but the
parallel search delays exploring the optimal branch. Alternately, super-linear
speedups are possible in case the parallel search happens on an optimal branch
that the sequential search does not explore until much later. In short, the
perturbation of the search order caused by adding processors makes it impossible
to \emph{predict} parallel performance.

These unusual performance characteristics make reproducible algorithmic
research into combinatorial search difficult: was it the new heuristic
that improved performance, or were we just lucky with the search ordering in
this instance? As the instances we wish to tackle become larger, parallelism is
becoming central to algorithmic research, and it is essential to be able to
reason about parallel performance.

This paper aims to develop a generic parallel branch and bound
search for distributed memory architectures (clusters).
Crucially, the objective is \emph{predictable parallel performance},
and the key to achieving this is careful control of the parallel search order.

The paper starts by illustrating performance anomalies with parallel
branch and bound by using a Maximum Clique graph search.
The paper then makes the following research contributions:

\begin{itemize}

\item To address search order related performance anomalies,
\cref{sec:motivatingExample} postulates three \emph{parallel search properties}
for replicable performance as follows.
\begin{description}
  \item[{\propA}:] Parallel runtime is never higher than sequential (one
    worker) runtime.

  \item[{\propB}:] Parallel runtime does not increase as the number of
    workers increases.

\item[{\propC}:] Parallel runtimes of repeated searches on the same parallel configuration have low variance.
\end{description}

\item We define a novel formal model for general parallel branch and
  bound backtracking search problems (BBM) that specifies both search
  order and parallel reduction (\cref{sec:formalModel}). We show the
  generality of BBM by using it to define three different benchmarks with a
  range of application areas: Maximum
  Clique (\cref{sec:formalModel}), 0/1 Knapsack
  (\cref{sec:formalKnapsack}) and Travelling Salesperson
  (\cref{sec:formalTSP}).

\item We define a new Generic Branch and Bound (GBB) search API that
  conforms to the BBM (\cref{sec:genericBBSearch}). The generality of
  the GBB is shown by using it to implement Maximum Clique
  (\cref{sec:motivatingExample})\footnote{This implementation being the first
  \textit{distributed-memory} parallel implementation of San Segundo's bit
  parallel Maximum Clique algorithm
  (BBMC)~\cite{segundo.matia.ea_improvedBBMC:2011}.}, 0/1 Knapsack
  (\cref{sec:GBBknapsack}) and Travelling Salesperson
  (\cref{sec:GBBTSP}).

\item To avoid the significant engineering effort required to produce
  a parallel implementation for each search algorithm we encapsulate
  the search behaviours as a pair of \emph{algorithmic
    skeletons}, that is, as generic polymorphic computation
  patterns~\cite{cole_skeletons:1991}, providing distributed memory
  implementations for the skeletons (\cref{sec:skeletons}).
  Both skeletons share the same API yet differ in how they schedule
  parallel tasks.
  The \emph{Unordered skeleton} relies on random work stealing, a tried and
  tested way to scale irregular task-parallel computations. In contrast, the
  \emph{Ordered skeleton} schedules tasks in an ordered fashion, closely
  following the sequential search order, so as to guarantee the parallel search
  properties.

\item We compare the sequential performance of the skeletons with a class
  leading hand tuned C++ search implementation, seeing slowdown factors of only
  between 1.9 and 6.2, and then assess whether the Ordered skeleton preserves
  the parallel search properties using 40 instances of the three benchmark
  searches on a cluster with 17 hosts and 200 workers (\cref{sec:evaluation}).
  The Ordered skeleton preserves all three properties and replicable results are
  achieved. The key results are summarised and discussed in
  \cref{sec:conclusion}.

\end{itemize}

\section{The Challenges of Parallel Branch and Bound Search}
\label{sec:motivatingExample}

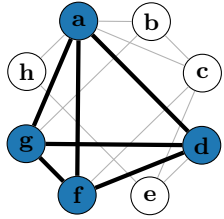
\begin{figure}
  \centering
    \begin{tikzpicture}
        \newcount \myc
        \foreach \n/\l in {1/a, 2/b, 3/c, 4/d, 5/e, 6/f, 7/g, 8/h}{
            \myc=\n \advance\myc by -1 \multiply\myc by -360 \divide\myc by 8 \advance\myc by 90
            \advance\myc by 22.5
            \ifthenelse{\n = 1 \OR \n = 4 \OR \n = 6 \OR \n = 7}{
                \node[selectedvertex] (N\n) at (\the\myc:1.25) {\l};
            }{
                \node[vertex] (N\n) at (\the\myc:1.25) {\l};
            }
        }

        \draw [edge] (N1) -- (N2);
        \draw [edge] (N1) -- (N3);
        \draw [edge] (N1) -- (N8);
        \draw [edge] (N2) -- (N3);
        \draw [edge] (N2) -- (N7);
        \draw [edge] (N3) -- (N5);
        \draw [edge] (N3) -- (N6);
        \draw [edge] (N4) -- (N5);
        \draw [edge] (N5) -- (N8);

        \draw [bedge] (N1) -- (N4);
        \draw [bedge] (N1) -- (N6);
        \draw [bedge] (N1) -- (N7);
        \draw [bedge] (N4) -- (N6);
        \draw [bedge] (N4) -- (N7);
        \draw [bedge] (N6) -- (N7);
    \end{tikzpicture}
    \caption{A graph, with its Maximum Clique $\{a, d, f, g\}$ shown.}
    \label{fig:maxcliqueExample}
\end{figure}

We start by considering a branch and bound search application, namely finding
the largest clique within a graph. The Maximum Clique problem appears as part of
many applications such as in bioinformatics ~\cite{Eblen:2011}, in
biochemistry~\cite{Butenko:2006,Konc:2007,Fukagawa:2011,Depolli:2013}, for
community detection \cite{Yan:2009}, for document clustering~\cite{Okubo:2006},
in computer vision, electrical engineering and communications~\cite{Bomze:1999},
for image comparison~\cite{SanSegundo:2010}, as an intermediate step in maximum
common subgraph and graph edit distance problems~\cite{McCreeshNPS:2016}, and
for controlling flying robots~\cite{Regula:2013}.

To illustrate the Maximum Clique problem we use the example graph
in~\cref{fig:maxcliqueExample}. In practice the graphs searched are
much larger, having hundreds or thousands of vertices. A clique within
a graph is a set of vertices where each vertex in the set is adjacent
to every other vertex in the set. For example,
in~\cref{fig:maxcliqueExample} the set $V = \{a,b,c\}$ is a clique as
all vertices are adjacent to one another. $\{a,b,h\}$ is not a clique
as there is no edge between $b$ and $h$. In the Maximum Clique problem
we wish to find a largest clique (there may be multiple of the same
size) in the graph.  Here we are interested in the \emph{exact}
solution requiring the full search space to be explored.

One approach to solving this problem would be to enumerate the power set of
vertices and check the clique property on each (ordering by largest set). While
this approach can work for smaller graphs, the number of combinations grows
exponentially with the number of nodes in the graph making it computationally
unfeasible for large graphs.

A better approach, particularly for larger graphs, is to only \emph{generate}
sets of vertices that maintain the clique property. This is the essence of the
\emph{branching} function. In the case of clique search, given any set of vertices, the set of candidate
choices is the set of vertices adjacent to all vertices in the current clique. Once there are no valid branching choices
left we can record the size of the clique and \emph{backtrack}.

Finally, we can go one step further with the addition of
\emph{bounding}. The idea of bounding is that a \emph{current best}
result, known as the incumbent, is maintained. For Maximum Clique this
corresponds to the size of the largest clique seen so far. At each step we
determine, using a bounding function, whether or not the current selection
of vertices and those remaining could possibly unseat the incumbent
and if it is impossible then backtracking can occur, reducing the size of
the search space. For the Maximum Clique example the maximum size,
given a current clique, may be estimated using a greedy colouring
algorithm: clearly, if we can colour the remaining vertices using $k$
colours (giving adjacent vertices different colours), then the current
clique cannot be grown by more than $k$ vertices.

Practical algorithms for the Maximum Clique problem were the subject of the
second DIMACS implementation challenge in 1993~\cite{Dimacs2}. In 2012,
Prosser~\cite{Prosser12} performed a computational study of exact maximum clique
algorithms, focusing on a series of algorithms using a colour
bound~\cite{TomitaS03,TomitaK07,TomitaSHTW10}, together with bit-parallel
variants~\cite{segundo.losada.ea_BBMC:2011,segundo.matia.ea_improvedBBMC:2011}
that represent adjacency lists using bitsets to gain increased performance via vectorised instructions.
Since then, ongoing research has looked at variations on these algorithms,
including reordering colour classes~\cite{McCreeshP14}, reusing
colourings~\cite{NikolaevBS15}, treating certain vertices
specially~\cite{SegundoT14}, and giving stronger (but more expensive) bounding
using rules based upon MaxSAT inference between colour
classes~\cite{LiFX13,LiJX15,SegundoNB15}. (A recent broader
review~\cite{DBLP:journals/eor/WuH15} considers both heuristic and exact
algorithms).

There have been three thread-parallel implementations of these
algorithms~\cite{McCreeshP13,DepolliKRTJ13,mccreesh.prosser_shapeOfTheSearchTree:2015},
the most recent makes use of detailed inside-search measurements to explain
\emph{why} parallelism works, and how to improve it. These studies have been
limited to multi-core systems. A fourth study~\cite{DBLP:conf/icde/XiangGA13}
attempted to use MapReduce on a similar algorithm, but only presented speedup
results on three of the standard DIMACS instances, all of which possess special
properties which make parallelism unusually
simple~\cite{mccreesh.prosser_shapeOfTheSearchTree:2015}.

For simplicity this paper uses a bit-parallel variant of the MCSa1
algorithm~\cite{Prosser12}, which is
BBMC~\cite{segundo.matia.ea_improvedBBMC:2011} with a simpler initial vertex
ordering. Crucially the algorithm is not straightforward, and that unlike the
na\"\i{}ve and overly simplistic algorithms typically used to demonstrate
skeletons, is both close to the state of the art and a realistic reflection of
modern practical algorithms.

\subsection{General Branch and Bound Search}
\label{sec:generalBBSearch}

Although we introduced branch and bound search in relation to the Maximum Clique
problem, it has much wider applications. It is commonly seen for global
optimisation problems~\cite{Morrison201679} where some property is either maximised or minimised
within a general search space. Two other examples where branch and bound search
may be used are given in~\cref{sec:applications-knapsack,sec:applications-tsp}.

The details and descriptions of these algorithms vary and we take a unifying
view using terminology from constraint programming. In general, a constraint
satisfaction or optimisation problem has a set of variables, each with a domain
of values. The goal is to give each variable one of the values from its domain,
whilst respecting all of a set of constraints that restrict certain
combinations of assignments. In the case of optimisation problems, we seek the
best legal assignment, as determined by some objective function.

Such problems may be solved by some kind of backtracking search.  Branch and
bound is a particular kind of backtracking search algorithm for optimisation
problems, where the best solution found so far (the \emph{incumbent}) is
remembered, and is used to prune portions of the search space based upon an
over-estimate (the \emph{bound} function) of the best possible solution within
an unexplored portion of the search space.

For example, when searching for a Maximum Clique (a subset of vertices, where
every vertex in the set is adjacent to every other in the set) in a graph, we
have a ``true or false'' variable for each vertex, with true meaning ``in the
clique''. We may branch on whether or not to include any given vertex, reject
any undecided vertices that are not adjacent to the vertex we just accepted,
and then bound the remaining search space using the colour bound mentioned
above.

In practice, selecting a good branching rule makes a huge difference. We must
select a variable, and then decide the value to assign it first. There are good
general principles for variable selection, but value ordering tends to be more
difficult in practice.

\subsection{Parallelisation and Search Anomalies}
\label{sec:searchAnomalies}

Search algorithms have strong dependencies: before we can evaluate a
subtree, we need to know the value of the incumbent from all the
preceding subtrees so we can determine if the bound can eliminate some
work.  Parallelism in these algorithms is \emph{speculative} as it
ignores the dependencies and creates tasks to explore subtrees in parallel.
This approach can lead to anomalous performance, and specifically.

\begin{enumerate}
\item When subtrees are explored in parallel
  some work may be wasted, since we might be exploring a
  subtree that would have been pruned in a sequential run by a
  stronger incumbent. As the parallel version is performing more work than the sequential version, its runtime may exceed that of the sequential version.
\item Conversely, it may be that a parallel task  finds a strong incumbent more quickly
  than in the sequential execution, leading to less work
  being done. In this case we observe superlinear speedups.
\item An absolute slowdown, where the parallel version runs
  exponentially slower than a sequential run. This can happen if introducing
  parallelism alters the search order, leading to it taking longer for a
  strong incumbent to be found.
\end{enumerate}

The theoretical conditions where these three conditions can occur are
well-understood~\cite{Lai:1984,Li:1985,Trienekens:1990,deBruin:1995}. In
particular, it is possible to guarantee that absolute slowdowns will never
happen, by requiring parallel search strategies to enforce certain
properties~\cite{deBruin:1995}.

\subsection{Implementation Challenges}
\label{sec:properties}

The most obvious complicating factor when parallelising a branch and bound
search tree is irregularity: it is extremely hard to decompose the problem
up-front to do static work allocation, since some subproblems are exponentially
more complicated than others.

To deal with irregular subproblems efficiently we require a form of dynamic load
balancing that can re-assign problems to cores as they become idle. A common
approach to dynamic load balancing in parallel search~\cite{olivier.huan.ea_UTS:2006} (and general parallelism) is
through \emph{work stealing}: we start with a sequential search, but allow
additional workers to ``steal'' portions of the search space and explore them in
parallel. Popular off-the-shelf work stealing systems commonly employ a
randomised stealing strategy, which has good theoretical
properties~\cite{blumofe.joerg.ea:1996}.

Surprisingly, though, irregularity is not the most
complex factor when parallelising these algorithms. Although non-linear speedups
are called \emph{anomalies} in the literature, anomalous behaviour is actually
extremely common when starting with strong sequential algorithms, to the extent
that if a linear speedup is reported, we should be suspicious as to why.
Although such behaviour is relatively uncommon with small numbers of cores,
e.g.\ four cores, our
experience~\cite{mccreesh.prosser_shapeOfTheSearchTree:2015} is that as we start
working in the 32 to 64 core range, anomalies often become the dominating factor
in the results. We expect that as core counts increase, such factors will become
even more important.

From an implementation perspective, anomalies cause serious complications, with
inconsistent and hard-to-understand speedup results being common.
Randomised work stealing schemes further complicate matters and recent
research~\cite{ChuSS09,MoisanQG14,mccreesh.prosser_shapeOfTheSearchTree:2015}
has demonstrated a connection between value-ordering heuristic
behaviour~\cite{HarveyG95} and parallel work splitting strategies that explains
anomalous behaviour. We now understand why randomised work stealing behaves so
erratically in practice in these settings: it interacts poorly with carefully
designed search order strategies~\cite{mccreesh.prosser_shapeOfTheSearchTree:2015}. For consistently strong
results, we cannot think of parallelism independently of the underlying
algorithm, and must instead use work stealing to explicitly offset the weakest
value ordering heuristic behaviour. For this reason, the best results for
parallel Maximum Clique algorithms currently come from handcrafted and complex
work distribution mechanisms requiring extremely intrusive modifications to
algorithms. It is not surprising that these implementations are currently
restricted to a single multi-core machine.

To conduct replicable parallel branch and bound research it is
essential to avoid these anomalies. To do so we propose that parallel
branch and bound search implementations should meet the following
properties\footnote{We are interested in parallel searches that meet or fail to meet these properties due to search order effects. We ignore resource related effects such as problem size being too small or massive oversubscription.}.

\begin{description}
  \item[{\propA}:] Parallel runtime is never higher than sequential (one
    worker) runtime.

  \item[{\propB}:] Parallel runtime does not increase as the number of
    workers increases.

\item[{\propC}:] Parallel runtimes of repeated searches on the same parallel configuration have low variance.
\end{description}

Engineering a parallel implementation that ensures these properties for each
search algorithm is non-trivial, and hence in~\cref{sec:skeletons} we
develop generic algorithmic branch and bound skeletons, which greatly
simplify the implementation of parallel searches.

\section{A Formal Model of Tree Traversals}
\label{sec:formalModel}


\newcommand{\ie}{i.\,e.\xspace}
\newcommand{\Ie}{I.\,e.,\xspace}
\newcommand{\eg}{e.\,g.,\xspace}
\newcommand{\Eg}{E.\,g.,\xspace}
\newcommand{\cf}{cf.\xspace}
\newcommand{\Cf}{Cf.\xspace}
\newcommand{\wrt}{w.\,r.\,t.\xspace}
\newcommand{\Wrt}{W.\,r.\,t.\xspace}
\newcommand{\wlg}{w.\,l.\,o.\,g.\xspace}
\newcommand{\Wlg}{W.\,l.\,o.\,g.\xspace}
\newcommand{\resp}{resp.\xspace}

\newcommand{\cons}{\mathord{:}}
\newcommand{\nil}{\mathord{[\,]}}

\newcommand{\nat}{\mathbb{N}}
\newcommand{\tup}[1]{\langle{#1}\rangle}

\newcommand{\set}[1]{\{{#1}\}}
\newcommand{\UNION}{\mathop{\bigcup}}
\newcommand{\union}{\mathbin{\cup}}
\newcommand{\INTER}{\mathop{\bigcap}}
\newcommand{\inter}{\mathbin{\cap}}

\newcommand{\dom}{\mathord{\mathit{dom}}}

\newcommand{\wmt}{\epsilon}                      
\newcommand{\wprfq}{\mathrel{\preceq}}           
\newcommand{\wprf}{\mathrel{\prec}}              
\newcommand{\wlexq}{\mathrel{\leq_{\mathrm{lex}}}}  
\newcommand{\wlex}{\mathrel{<_{\mathrm{lex}}}}      

\newcommand{\qleq}{\mathrel{\sqsubseteq}}        
\newcommand{\qgeq}{\mathrel{\sqsupseteq}}
\newcommand{\qlt}{\mathrel{\sqsubset}}           
\newcommand{\qgt}{\mathrel{\sqsupset}}

This section formalises parallel backtracking traversal of
search trees with pruning, modeling the behaviour of a multi-threaded
branch-and-bound algorithm in the reduction style of operational semantics.
This formal model, for brevity refered to as \emph{BBM}, admits reasoning
about the effects of parallel reductions, in particular how parallelism
affects the potential to prune the search space.

Reduction-based operational semantics of algorithmic skeletons has
been studied previously~\cite{AldinucciD:COMLAN:2007} for standard
stateless skeletons like pipelines and maps.  BBM does not fit this
stateless framework since branch and bound skeletons maintain state in
the form a globally shared incumbent.
There are several theoretical analyses of parallel branch and bound
search~\cite{gendron.crainic_parallelBranchAndBoundSurvey:1994}, often specific
to a particular search algorithm. BBM is novel in encoding generic branch and
bound searches as a set of parallel reduction rules.

\subsection{Modelling Trees and Tree Traversals}
\label{sec:tree}

In practice, search trees are implicit. They are not materialised
as data structures in memory but traversed in a specific order,
for instance depth-first.
In contrast, for the purpose of this formalisation we assume the
search tree is fully materialised.
This is not a restriction as the search tree is typically generated
by a tree generator. In practice, the tree generator is interleaved
with the tree traversal avoiding the need to materialise
the search tree in memory.

We  formalise trees as prefix-closed sets of words.
To this end, we introduce some notation.
Let $X$ be a non-empty set.
By $2^X$, we denote the power set of $X$.
We denote the set of finite words over alphabet $X$ by $X^*$,
and the empty word by $\wmt$.
We write $|w|$ to denote the length of a word $w \in X^*$.

We denote the prefix order on $X^*$ by $\wprfq$.
We write $(w \wprfq)$ to denote the principal filter for $w \in X^*$,
that is, $(w \wprfq) = \set{v \in X^* \mid w \wprfq v}$.

By $\wlexq$, we denote the lexicographic extension of the natural order
$\leq$ on $\nat$ to $\nat^*$.
Note that $\wlexq$ is an extension of the prefix order $\wprfq$, that is,
being prefix-ordered implies being ordered lexicographically
on words in $\nat^*$.

\paragraph{Trees}

A \emph{tree $T$ over alphabet $X$} is a non-empty subset of $X^*$
such that there is a least (\wrt the prefix-order) element $u \in T$,
and $T$ is prefix-closed above $u$.
Formally, $T$ is prefix-closed above $u$ if for all $v, w \in X^*$,
$u \wprfq v \wprfq w$ and $w \in T$ implies $v \in T$.
When $X$ and $u$ are understood, we will simply call $T$ a \emph{tree}.
We call the elements of $T$ \emph{vertices}.
We call the least element $u \in T$ the \emph{root}; and we call
$v \in T$ a \emph{leaf} if it is maximal \wrt the prefix order, that is,
if there is no $w \in T$ with $v \wprf w$.
We call two distinct vertices $w, w' \in T$ \emph{siblings} if there are
$v \in X^*$ and $a, a' \in X$ such that $w = v a$ and $w = v a'$.

\begin{figure}
\includegraphics[width=\linewidth]{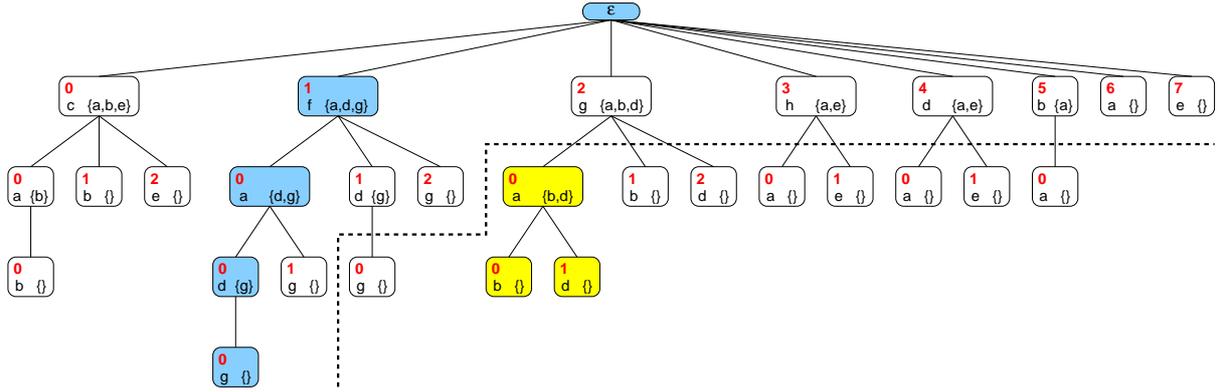}
\caption{Depiction of an ordered tree.
         The path in blue identifies the leaf $1000$;
         the vertices in yellow make up the tree segment rooted at $20$.
         The vertices below the dashed line are cut off
         by a sequential branch and bound traversal.}
\label{fig:treeXmpl}
\end{figure}

\Cref{fig:treeXmpl} depicts an example tree over the natural numbers.
That is, each vertex corresponds to the unique sequence of red numbers
from the root $\wmt$.
For example, the blue leaf is vertex $1000$, whereas the yellow non-leaf
is vertex $20$.

We call a function $g : X^* \to 2^X$ a \emph{tree generator}.
Given such a tree generator $g$, we define $t_g$ as the smallest subset
of $X^*$ that contains $\wmt$ and is closed under $g$ in the following sense:
For all $u \in t_g$ and all $a \in g(u)$, $u a \in t_g$.
Clearly, $t_g$ is a tree with root $\wmt$,
\emph{the} tree generated by $g$.

\paragraph{Subtrees and segments}

Let $T$ be a tree.
A subset $S$ of vertices of $T$ is a \emph{subtree} of $T$
if $S$ is a tree.
Given a vertex $u \in T$, we call the greatest (with respect to set inclusion)
subtree $S$ of $T$ with root $u$ the \emph{segment} of $T$ rooted at $u$.
The yellow vertices in \cref{fig:treeXmpl} depict the
segment $\set{20, 200, 201}$, rooted at vertex $20$.

Two segments of $T$ are \emph{overlapping} if they intersect non-trivially,
in which case one is contained in the other.
A set of segments \emph{cover} the tree $T$
if the prefix-closure of their union equals $T$.
That is, if for each $u \in T$ there is a segment $S$ and $v \in S$
such that $u \wprfq v$.

\paragraph{Ordered trees}
Trees as defined above capture the parent-child relation
(via the prefix order on words) but do not impose any order on siblings.
Yet, many tree traversals rely on a specific order on siblings.
To be able to express such an order, we generalise the notion of trees
to \emph{ordered} trees.
We do so by labeling trees over the natural numbers, using the usual order
of the naturals (or rather, its lexicographic extension to words)
to order siblings.

Formally, an \emph{ordered tree $\lambda$ over $X$} is a function
$\lambda : \dom(\lambda) \to X^*$ such that
\begin{itemize}
\item $\dom(\lambda)$ is a tree over $\nat$,
\item the image of $\lambda$ is a tree over $X$, and
\item $\lambda$ is an order isomorphism between the two trees,
      both ordered by the prefix order $\wprfq$.
\end{itemize}
Since $\lambda$ is an isomorphism of the prefix order
the lengths of the words $u$ and $\lambda(u)$ coincide
for all $u \in \dom(\lambda)$.
In an abuse of notation, we write $\lambda$ to denote both the ordered
tree (\ie the function from $\dom(\lambda)$ to $X^*$) as well as the
corresponding tree over $X$ (\ie the image of the function $\lambda$).
When $X$ is understood, we will simply call $\lambda$ an \emph{(ordered) tree}.
To avoid confusion, we will call the elements of $\lambda$ \emph{vertices},
and the elements of $\dom(\lambda)$ \emph{positions}.

\Cref{fig:treeXmpl} shows an example  ordered tree where each node corresponds to the string of red numbers
from the root to that node, i.e.\ a tree over $\nat$.
The figure also depicts an ordered tree $\lambda$ over the alphabet
$X = \set{\textsf{a},\dots,\textsf{h}}$, where $\lambda$ maps each position
to the string of black letters from the root to the corresponding node.
For instance $\lambda$ maps position $1000$ to the string
$\textsf{fadg}$ which happens to represent the maximum clique
of the graph in~\cref{fig:maxcliqueExample}.

As $\lambda$ is an order isomorphism the lexicographic ordering on
$\dom(\lambda)$ carries over to the tree $\lambda$.  That is, we
define for all $u, v \in \dom(\lambda)$, $\lambda(u) \wlexq
\lambda(v)$ if and only if $u \wlexq v$, and $\wlexq$ becomes a total
ordering on $\lambda$.

We call a function $g : X^* \to X^*$ an \emph{ordered tree generator}
if all images of $g$ are isograms, \ie have no repeating letters.
Given an ordered tree generator $g$,
we define $\lambda_g : \dom(\lambda_g) \to X^*$ as the function with
smallest domain such that
\begin{itemize}
\item $\dom(\lambda_g)$ is a tree over $\nat$,
\item $\lambda_g(\wmt) = \wmt$, and
\item $\lambda_g$ is closed under $g$ in the following sense:
      For all positions $u \in \dom(\lambda_g)$ and corresponding
      vertices $v = \lambda_g(u)$,
      if $g(v) = a_0 a_1 \dots a_{n-1}$ and $i < n$
      then $u i$ is a position in $\dom(\lambda_g)$ and
      $\lambda_g(u i) = v a_i$.
\end{itemize}
By construction $\lambda_g$ is an order isomorphim as images of $g$
are isograms, hence $\lambda_g$ is an ordered tree,
\emph{the} ordered tree generated by $g$.

\paragraph{Example: Tree generators for clique problems}

Let $G = \tup{V,E}$ be an undirected graph.
Given a vertex $u \in V$, we denote its set of neighbours by $E(u)$.

We define $g : V^* \to 2^V$ by $g(u_1 \dots u_m) = \set{v \in V \mid
\forall i : v \neq u_i \wedge u_i \in E(v)}$.  Clearly, $g$ is a
generator for the tree $t_g$ over the alphabet $X = V$, enumerating
all cliques of $G$.  However, $t_g$ enumerates cliques as strings
rather than sets and hence every clique of size $k$ will be enumerated
$k!$ times.


To avoid enumerating the same clique multiple times, we need to generate
an ordered tree where siblings ``to the right'' avoid vertices that have
already been chosen ``on the left''.
We construct an ordered tree over the alphabet $X = V \times 2^V$,
where the first component is the latest vertex added to the current clique and
the second component is a set of candidate vertices that may extend
the current clique. The candidate vertices are incident to all vertices
of the current clique, but do not necessarily form a clique themselves.
We define the ordered tree generator $h : X^* \to X^*$ by
$h(\tup{u_1,U_1}\dots\tup{u_m,U_m}) = \tup{v_1,V_1}\dots\tup{v_n,V_n}$ such that
\begin{itemize}
\item the $v_i$ enumerate the set $U$, and
\item the $V_i = (U \setminus \set{v_1,\dots,v_{i-1}}) \inter E(v_i)$
\end{itemize}
where $U = U_m$ if $m > 0$, and $U = V$ otherwise.
Typically, the $\tup{v_i,V_i}$ are ordered such that the size
of $V_i$ decreases as $i$ increases; this order is beneficial for
sequential branch and bound traversals.

Clearly, $h$ is an ordered generator for an ordered tree enumerating all cliques
of $G$ \emph{exactly once} (ignoring the second component of the alphabet).
\Cref{fig:treeXmpl} shows a tree generated by $h$ for the graph
from~\cref{fig:maxcliqueExample}.

\subsection{Maximising Tree Traversals}
\label{sec:trav}

The trees defined above materialise the search space and order
traversals.  What is needed for modeling branch-and-bound is an
\emph{objective function} to be computed during traversal and that the
search aims to maximise.

Let $Y$ be a set with a total quasi-order $\qleq$, that is
$\qleq$ is a reflexive and transitive, but not necessarily anti-symmetric,
total binary relation on $Y$.

Given a tree $T$ over $X$ and an \emph{objective} function $f : X^* \to Y$,
the goal is to maximise $f$ over $T$, \ie to find some $u \in T$ such that
$f(u) \qgeq f(v)$ for all $v \in T$.
The objective function is required to be monotonic \wrt the prefix order,
that is for all $u, u' \in X^*$, if $u \wprfq u'$ then $f(u) \qleq f(u')$.
By monotonicity $f(\wmt)$ is a minimal element of the image of $f$.

So far, we have modeled maximising tree search.
To model branch-and-bound we introduce one additional refinement:
A predicate $p$ for \emph{pruning} subtrees that cannot improve the incumbent.
More precisely, the \emph{pruning} predicate $p : Y \times X^* \to \set{0,1}$
is a function mapping the incumbent (\ie the maximal value of $f$ seen so far)
and the current vertex to $1$ (for \emph{prune}) or $0$ (for \emph{explore}).
The pruning predicate must satisfy the following monotonicity and
compatibility conditions:
\begin{enumerate}
\item\label[condition]{enum:prune:i}
  For all $y \in Y$ and $u, u' \in X^*$,
  if $u \wprfq u'$ then $p(y,u) \leq p(y,u')$.
\item\label[condition]{enum:prune:ii}
  For all $y, y' \in Y$ and $u \in X^*$,
  if $y \qleq y'$ then $p(y,u) \leq p(y',u)$.
\item\label[condition]{enum:prune:iii}
  For all $y \in Y$ and $u \in X^*$,
  if $p(y,u) = 1$ then $f(u) \qleq y$.
\end{enumerate}
\Cref{enum:prune:i} implies that all descendents $u'$
of a pruned vertex $u$ are also pruned.
\Cref{enum:prune:ii} implies a vertex pruned by incumbent $y$
is also pruned by any stronger incumbent $y'$.
Finally, \cref{enum:prune:iii} states the correctness of pruning
\wrt maximising the objective function:
Vertex $u$ is pruned by incumbent $y$ only if $f(u)$ does not beat $y$.

How exactly pruning will interact with the tree traversal will be detailed
in the next section.
Note that pruning is an \emph{optimisation} and must not be used to
constrain the search space. That is, the result of the tree traversal
must be independent of the pruning predicate.
In particular, the trivial pruning predicate that always returns $0$
(and hence prunes nothing) is a legal predicate.

\paragraph{Example: Objective function and pruning predicate for clique problems}

For maximum clique, we set $Y = \nat$, and the quasi-order $\qleq$ is
the natural order $\leq$.
We define the objective function
$f : X^* \to Y$ by $f(w) = |w|$.
That is, maximising $f$ means finding cliques of maximum size.
We define the pruning predicate $p : Y \times X^* \to \set{0,1}$ by
\begin{equation*}
p(l, \tup{\_,U_1}\dots\tup{\_,U_m}) = \left\{\begin{array}{ll}
1 & \text{if}~m > 0~\text{and}~m + |U_m| \leq l \\
0 & \text{otherwise}
\end{array}\right.
\end{equation*}
That is, pruning decisions rest on the size of the current clique, $m$,
and the size of the set of remaining candidate vertices $U_m$;
vertices will be pruned if adding these two sizes does not exceed
the current bound $l$.\footnote{More accurate pruning can be achieved
by replacing the size of $U_m$ with the size of the maximum clique
of the subgraph induced by $U_m$; greedily colouring this subgraph makes for an
efficent approximation of maximum clique size.}

\subsection{Modelling Multi-threaded Tree Traversals}
\label{sec:model}

For this section, we fix an ordered tree $\lambda$ over $X$, which we
will traverse according to the order $\wlexq$.
We also fix an objective function $f : X^* \to Y$,
and a pruning predicate $p : Y \times X^* \to \set{0,1}$,
where $Y$ is a set with a total quasi-order $\qleq$.
Finally, we fix a set $\mathit{SEG}$ of pairwise non-overlapping
tree segments that cover the tree $\lambda$;
we call each segment $S \in \mathit{SEG}$ a \emph{task}.

\paragraph{State}

Let $n \geq 1$ be the number of threads.
The \emph{state} of a backtracking tree traversal is a $(n+2)$-tuple of the form
$\sigma = \tup{x, \mathit{Tasks}, \theta_1, \dots, \theta_n}$, where
\begin{itemize}
\item $x \in \lambda$ is the \emph{incumbent},
      \ie the vertex that currently maximises $f$,
\item $\mathit{Tasks} \in \mathit{SEG}^*$ is a queue of pending tasks, and
\item $\theta_i$ is the state of the $i$-th thread, where $\theta_i = \bot$
      if the $i$-th thread is idle, or $\theta_i = \tup{S_i,v_i}$ if
      $S_i \in \mathit{SEG}$ is the $i$-th thread's current task
      and $v_i \in S_i$ the currently explored vertex of that task.
\end{itemize}
We use Haskell list notation for the task queue $\mathit{Tasks}$.
That is, $\nil$ denotes the empty queue, and $S\cons\mathit{Tasks}$ denotes
a non-empty queue with head $S  \in \mathit{SEG}$.

The \emph{initial state} is $\tup{\wmt, \mathit{Tasks}, \bot, \dots, \bot}$,
where the list $\mathit{Tasks}$ enumerates all tasks in $\mathit{SEG}$, in an
arbitrary but fixed order.
A \emph{final state} is of the form $\tup{x, \nil, \bot, \dots, \bot}$.

\paragraph{Reductions}

\begin{figure}
\begin{equation*}
\renewcommand{\arraystretch}{3.0}
\begin{array}{l@{~}l}
(\text{strengthen}_i) &
\displaystyle\frac%
{f(x) \qlt f(v_i)}%
{\tup{x, \mathit{Tasks},\dots,\tup{S_i,v_i},\dots}
 \to
 \tup{v_i, \mathit{Tasks},\dots,\tup{S_i,v_i},\dots}}
\\
(\text{schedule}_i) &
\displaystyle\frac%
{v_i = \text{root of}~S
 \qquad
 p(f(x),v_i) = 0}%
{\tup{x, S\cons\mathit{Tasks},\dots,\bot,\dots}
 \to
 \tup{x, \mathit{Tasks},\dots,\tup{S,v_i},\dots}}
\\
(\text{prune}_i) &
\displaystyle\frac%
{v_i = \text{root of}~S
 \qquad
 p(f(x),v_i) = 1}%
{\tup{x, S\cons\mathit{Tasks},\dots,\bot,\dots}
 \to
 \tup{x, \mathit{Tasks},\dots,\bot,\dots}}
\\
(\text{advance}_i) &
\displaystyle\frac%
{v_i' \in S_i
 \qquad
 v_i \wlex v_i'
 \qquad
 p(f(x), v_i') = 0
 \qquad
 \forall v \in S_i\,
   \bigl(v_i \wlex v \wlex v_i' \Rightarrow p(f(x),v) = 1\bigr)}%
{\tup{x, \mathit{Tasks}, \dots,\tup{S_i,v_i},\dots}
 \to
 \tup{x, \mathit{Tasks}, \dots,\tup{S_i,v_i'},\dots}}
\\
(\text{terminate}_i) &
\displaystyle\frac%
{\forall v \in S_i\,\bigl(v_i \wlex v \Rightarrow p(f(x),v) = 1\bigr)}%
{\tup{x, \mathit{Tasks}, \dots,\tup{S_i,v_i},\dots}
 \to
 \tup{x, \mathit{Tasks}, \dots,\bot,\dots}}
\end{array}
\end{equation*}
\caption{Reduction rules.}
\label{fig:rules}
\end{figure}

The reduction rules in \cref{fig:rules} define a binary relation $\to$
on states.
Each rule carries a subscript indicating which thread it is operating on.
Rule $(\text{strengthen}_i)$ is applicable if the $i$-th thread is not idle and
its current vertex $v_i$ beats the incumbent on $f$.
Of the remaining four rules exactly one will be applicable to the $i$-th thread
(unless a final state is reached).

Rules $(\text{schedule}_i)$ and $(\text{prune}_i)$ apply if the $i$-th thread
is idle and the task queue is non-empty.
Which of the two rules applies depends on whether the root vertex $v_i$
of the head task $S$ in the queue is to be pruned or not.
If not, $S$ becomes the $i$-th thread's current task and $v_i$ the current
vertex, otherwise task $S$ is pruned and the $i$-th thread remains idle.

Rules $(\text{advance}_i)$ and $(\text{terminate}_i)$ apply if the $i$-th thread
is not idle.
Which of the two rules applies depends on whether all vertices of the current
task $S_i$ beyond the current vertex $v_i$ (in the lexicographic order $\wlex$)
are to be pruned according to predicate $p$.
If so, the $i$-th thread terminates the current task and becomes idle,
otherwise the thread advances to the next vertex $v_i'$ that is not pruned.

It is easy to see that no rule is applicable if and only if all threads are
idle and the task queue is empty, that is, iff a final state is reached.

\paragraph{Admissible reductions}

The reduction rules in~\cref{fig:rules} do not specify an ordering
on the rules nor stipulate any restriction on the relative speed of execution
of different threads.
However, applying the rules in just any order is too liberal.
In particular, not selecting rule $(\text{strengthen}_i)$ when the incumbent
could in fact be strengthened may result in missing the maximum.
To avoid this, rule $(\text{strengthen}_i)$ must be prioritised as follows.

We call a reduction $\sigma \to \sigma'$ \emph{inadmissible}
if it uses rule $(\text{advance}_i)$ or $(\text{terminate}_i)$
even though rule $(\text{strengthen}_i)$ was applicable in state $\sigma$.
A reduction is \emph{admissible} if it is not inadmissible.
Admissible reductions prioritise rule $(\text{strengthen}_i)$ over
rules $(\text{advance}_i)$ and $(\text{terminate}_i)$.

By induction on the length of the reduction sequence, one can show that
an incumbent $x$ maximises the objective function $f$ over the ordered
tree $\lambda$ whenever $\tup{x, \nil, \bot, \dots, \bot}$ is a final state
reachable from the initial state $\tup{\wmt, \mathit{Tasks}, \bot, \dots, \bot}$
by a sequence of admissible reductions.

We point out that final states are generally not unique.
For instance, a graph may contain several different cliques of maximum size,
and a parallel maxclique search may non-deterministically return any
of these maximum cliques.
Therefore the reduction relation cannot be confluent.

\paragraph{Example: Reductions for maxclique}

Consider the tree in~\cref{fig:treeXmpl} encoding the graph
in~\cref{fig:maxcliqueExample}. Let $\mathit{Tasks} =
[S_0,S_1,S_2,S_3,S_4,S_5,S_6,S_7]$ be a queue of tasks such that $S_i$ is the
segment rooted at vertex $i$; for example the segment $S_2$ is determined by the
set of positions $\set{2, 20, 200, 201, 21, 22}$. Clearly, the $S_i$ are
pairwise non-overlapping and cover the whole tree. Below, we consider a sample
reduction with three threads (with IDs 1 to 3) following a strict round-robin
thread scheduling policy, except for selecting the strengthening rule
\emph{eagerly} (that is, as soon as it is applicable).
For convenience, we display the reduction rule used in the left-most column and
index the reduction arrow with the number of reductions.

\begin{figure}[h!]
\begin{equation*}
\begin{array}{rcl|l|l}
& & \tup{\wmt, [S_0,S_1,S_2,S_3,S_4,S_5,S_6,S_7], \bot, \bot, \bot} &
\text{pruned} & \text{cut off} \\
(\text{schedule}_1) & \to_{1} &
\tup{\wmt, [S_1,S_2,S_3,S_4,S_5,S_6,S_7], \tup{S_0,0}, \bot, \bot} & & \\
(\text{strengthen}_1) & \to_{2} &
\tup{0, [S_1,S_2,S_3,S_4,S_5,S_6,S_7], \tup{S_0,0}, \bot, \bot} & & \\
(\text{schedule}_2) & \to_{3} &
\tup{0, [S_2,S_3,S_4,S_5,S_6,S_7], \tup{S_0,0}, \tup{S_1,1}, \bot} & & \\
(\text{schedule}_3) & \to_{4} &
\tup{0, [S_3,S_4,S_5,S_6,S_7], \tup{S_0,0}, \tup{S_1,1}, \tup{S_2,2}} & & \\
(\text{advance}_1) & \to_{5} &
\tup{0, [S_3,S_4,S_5,S_6,S_7], \tup{S_0,00}, \tup{S_1,1}, \tup{S_2,2}} & & \\
(\text{strengthen}_1) & \to_{6} &
\tup{00, [S_3,S_4,S_5,S_6,S_7], \tup{S_0,00}, \tup{S_1,1}, \tup{S_2,2}} & & \\
(\text{advance}_2) & \to_{7} &
\tup{00, [S_3,S_4,S_5,S_6,S_7], \tup{S_0,00}, \tup{S_1,10}, \tup{S_2,2}} & & \\
(\text{advance}_3) & \to_{8} &
\tup{00, [S_3,S_4,S_5,S_6,S_7], \tup{S_0,00}, \tup{S_1,10}, \tup{S_2,20}} & & \\
(\text{advance}_1) & \to_{9} &
\tup{00, [S_3,S_4,S_5,S_6,S_7], \tup{S_0,000}, \tup{S_1,10}, \tup{S_2,20}} & & \\
(\text{strengthen}_1) & \to_{10} &
\tup{000, [S_3,S_4,S_5,S_6,S_7], \tup{S_0,000}, \tup{S_1,10}, \tup{S_2,20}} & &\\
(\text{advance}_2) & \to_{11} &
\tup{000, [S_3,S_4,S_5,S_6,S_7], \tup{S_0,000}, \tup{S_1,100}, \tup{S_2,20}} &&\\
(\text{terminate}_3) & \to_{12} &
\tup{000, [S_3,S_4,S_5,S_6,S_7], \tup{S_0,000}, \tup{S_1,100}, \bot} &
200, 201, 21, 22 & \\ 
(\text{terminate}_1) & \to_{13} &
\tup{000, [S_3,S_4,S_5,S_6,S_7], \bot, \tup{S_1,100}, \bot} &
01, 02 & \\ 
(\text{advance}_2) & \to_{14} &
\tup{000, [S_3,S_4,S_5,S_6,S_7], \bot, \tup{S_1,1000}, \bot} & & \\
(\text{strengthen}_2) & \to_{15} &
\tup{1000, [S_3,S_4,S_5,S_6,S_7], \bot, \tup{S_1,1000}, \bot} & & \\
(\text{prune}_3) & \to_{16} &
\tup{1000, [S_4,S_5,S_6,S_7], \bot, \tup{S_1,1000}, \bot} &
3 & 30, 31 \\ 
(\text{prune}_1) & \to_{17} &
\tup{1000, [S_5,S_6,S_7], \bot, \tup{S_1,1000}, \bot} &
4 & 40, 41 \\ 
(\text{terminate}_2) & \to_{18} &
\tup{1000, [S_5,S_6,S_7], \bot, \bot, \bot} &
101, 11, 12 & 110 \\ 
(\text{prune}_3) & \to_{19} &
\tup{1000, [S_6,S_7], \bot, \bot, \bot} &
5 & 50 \\ 
(\text{prune}_1) & \to_{20} &
\tup{1000, [S_7], \bot, \bot, \bot} &
6 & \\ 
(\text{prune}_2) & \to_{21} &
\tup{1000, [], \bot, \bot, \bot} &
7 & \\ 
\end{array}
\end{equation*}
\caption{Example apply reduction rules}
\label{fig:ExampleReduction}
\end{figure}

We observe that up to reduction 11, the three threads traverse the search
tree segments $S_0$, $S_1$ and $S_2$ in parallel.
From reduction 12 onwards, the incumbent is strong enough to enable pruning
according to the heuristic, i.e.\ prune if size of current clique plus
number of candidates does not beat size of the incumbent.
Column \emph{pruned} lists the positions of the search tree where traversal
stopped due to pruning; column \emph{cut off} list the positions that were
never reached due to pruning.
The reduction illustrates that parallel traversals potentially do more work
than sequential ones in the sense that fewer positions are cut off.
Concretely, thread 3 traverses segment $S_2$ because the incumbent
is too weak; a sequential traversal would have entered $S_2$ with the
final incumbent and pruned immediately, as indicated by the dashed line
in~\cref{fig:treeXmpl}.
The reduction also illustrates that parallelism may reduce
runtime: a sequential traversal would explore first $S_0$ and then $S_1$,
whereas thread 2 locates the maximum clique in $S_1$ without traversing
$S_0$ first.


\section{Generic Branch and Bound Search}
\label{sec:genericBBSearch}

This section uses the model in~\cref{sec:formalModel} as the
basis of a Generic Branch and Bound (GBB) API for specifying  search
problems.  The GBB API makes extensive use of higher-order functions,
i.e.\ functions that take functions as arguments, and hence is suitable
for parallel implementation in the form of skeletons (\cref{sec:skeletons}).

We introduce each of the GBB API functions, give
their types and show an example of how to use them in a simple
implementation of the Maximum Clique problem
(\cref{sec:motivatingExample}). Later sections show that the
API is general enough to encode other branch and bound applications
(\cref{sec:applications-knapsack,sec:applications-tsp}).

We start by considering the key types and functions required to
specify a general branch and bound search. The API functions and types
are specified in Haskell~\cite{haskell} in~\cref{lst:BaBAPI}.

\begin{haskellcode}{Generic Branch and Bound (GBB) search API}{lst:BaBAPI}
-- application dependent types
type Space            -- data (e.g. graph) relevant to the problem
type PartialSolution  -- partial solution of the problem
type Candidates       -- set of candidates for extending the partial solution
type Bound            -- "size" of the partial solution; instance of Ord

-- type of nodes making up the search tree
type Node = (PartialSolution, Candidates, Bound)

-- generates a list of candidate Nodes for extending the search tree by
-- extending the PartialSolution of the given Node with each of the Candidates
orderedGenerator :: Space -> Node -> [Node]

-- Returns an upper bound on the size of any solution that could result
-- from extending the given Node's PartialSolution with any of the given
-- Candidates
pruningHeuristic :: Space -> Node -> Bound
\end{haskellcode}

\subsection{Types}

The fundamental type for a search is a \emph{Node} that represents a single
position within a search tree (for example in~\cref{fig:treeXmpl} each box
represents a node). This notion of a node differs slightly from the BBM where a
single type, \(X^*\), is used to uniquely identify a particular tree node by the
branches leading to it. For an efficient implementation, rather than store an
encoding of the branch through the tree, the node type uses the partial solution to encode
the branch history and the candidate set to encode potential next steps in the
branch. The current bound is maintained for efficiency reasons but could
alternatively be calculated from the current solution as in the BBM.

The abstract types are described below, and~\cref{tab:typeMappings}
shows how the abstract types map to implementation specific types for
Maximum Clique (\cref{sec:motivatingExample}), knapsack
(\cref{sec:applications-knapsack}) and travelling salesperson
(\cref{sec:applications-tsp}) searches.

\begin{description}

\item[Space:] Represents the domain specific structure to be searched.

\item[Solution:] Represents the current (partial) solution at this
  node. The solution is an application specific representation of a branch
  within the tree and encodes the history of the search.

\item[Candidates:] Represents the set of candidates that may still be added to
  the solution to extend the search by a single step. This may be used to encode
  implementation specific details such as no non-adjacent nodes in a maximum
  clique search, or simply ensure that no variable is chosen twice.

  It is not required that the type of the candidates matches the type the search
  space. This enables implementation-specific optimisations such as the bitset
  encoding found in the BBMC algorithm (\cref{sec:eval-maxclique}).

\item[Bound:] Represents the bound computed from the current solution. There
  must be an ordering on bounds, for example as provided by Haskell's Ord
  typeclass instance~\cite{haskellOrdTypeClass} to allow a maximising tree
  traversal to be performed implicitly using the type.

\item[Node:] Represents a position within the search space. For efficiency it
  caches the current bound, current solution and candidates for expansion.

\end{description}

\begin{table}[h]
  \centering
  \begin{tabular}{@{}l l l l@{}}
    \toprule
    \textbf{Abstract Type} & \textbf{Maximum Clique} & \textbf{Knapsack} & \textbf{TSP}\\
    \midrule
    \textbf{Space}    & Graph & List of all Items & Distance Matrix \\
    \textbf{Solution} & List of chosen vertices & List of chosen items & Current (partial) Tour \\
    \textbf{Candidates} & Vertices adjacent to all solution vertices & All remaining items & All remaining cities \\
    \textbf{Bound}    & Size of the current chosen vertices list & Current profit of items & Current tour length \\
    \bottomrule
  \end{tabular}
  \caption{Abstract to concrete type mappings}
  \label{tab:typeMappings}
\end{table}

\subsubsection{Function Usage}

It is perhaps surprising that the application specific aspects of a
branch and bound search can be both precisely specified, and
efficiently implemented, with just two functions. The GBB API functions
rely on the implicit ordering on the bound type, but could easily be
extended to take an ordering function as an argument.

\begin{description}

\item[orderedGenerator:] generates the set of candidate child nodes from a node
in the space. Search heuristics can be encoded by ordering the child nodes in a
list. The search ordering may use these heuristics to provide simple in-order
tree traversal or more elaborate heuristics such as depth based discrepancy search
(\cref{sec:eval-maxclique}).

\item[pruningHeuristic:] returns a speculative \emph{best} possible bound for
the current node. If this bound cannot unseat the global maximum then early
backtracking should occur as it is impossible for child nodes to beat
the current incumbent.

\end{description}

These functions correspond to the \emph{branching} and \emph{bounding} functions
respectively. We chose to call them \emph{orderedGenerator} and \emph{pruningHeuristic} to
highlight their purposes: to generate the next steps in the search and to
determine if pruning should occur.

\Cref{lst:maxclique-impl} shows instances of these GBB functions that
encode a simple, \emph{IntSet} based, version of the Maximum Clique
search. The \emph{orderedGenerator} builds a set of candidate nodes
based on a greedy graph colouring algorithm (colourOrder). The
colourings provide a heuristic ordering and, by storing them alongside
the solution's vertices, allow effective bounding to be performed.
Candidates only include vertices that are adjacent to every vertex
already in the clique. The \emph{pruningHeuristic} checks if the
number of vertices in the current clique and potential colourings can
possibly unseat the incumbent. See~\cref{sec:eval-maxclique} for
instances of the GBB API that use a more realistic bitset
encoding~\cite{segundo.losada.ea_BBMC:2011,
  segundo.matia.ea_improvedBBMC:2011}.

\begin{haskellcode}{Maximum Clique problem using the GBB API}{lst:maxclique-impl}
type Vertex = Int
type VertexSet = IntSet
type Colour = Int

type Space  = Graph
type PartialSolution = ([Vertex], Colour)
type Candidates = VertexSet
type Bound = Int
type Node  = (PartialSolution, Bound, Candidates)

colourOrder :: Graph -> VertexSet -> [(Vertex, Colour)]
colourOrder = -- defined elsewhere

-- Reduce a list to a value of type b
foldl :: (b -> a -> b) -> b -> [a] -> b
foldl f accumulator []     = accumulator
foldl f accumulator (x:xs) = foldl (f accumulator x) xs

orderedGenerator :: Graph -> Node -> [Node]
orderedGenerator graph ((clique, colour), candidates, size) =
  let choices = colourOrder graph candidates
  in fst (foldl buildNodes ([], candidates) choices)
  where
    buildNodes :: ([Node], VertexSet) -> (Vertex, Colour) -> ([Node], VertexSet)
    buildNnodes (nodes, candidates) (v, colour) = let
      newClique = (v : clique, colour - 1)
      newSize   = size + 1
      newCandidates  = VertexSet.intersection candidates (adjVertices graph v)
      -- We delete v from candidates to avoid generating duplicate solutions
      -- from any vertex "to-the-left" of the current
      in (nodes ++ [(newClique, newSize, newCandidates)], VertexSet.delete v candidates)

pruningHeuristic :: Graph -> Node -> Bound
pruningHeuristic g ((clique, colour), bnd, candidates) = bnd + colour
\end{haskellcode}

\subsection{General branch and bound Search Algorithm}

The essence of a branch and bound search is a recursive function for traversing
the nodes of the search space. \Cref{alg:parallelCoordination} shows the
function expressed in terms of the GBB API~(\cref{lst:BaBAPI}) where we assume
that the incumbent and associated bound are read and written by function calls
rather than being explicitly passed as arguments and returned as a result. Hence
the final solution is read from the global accessor function instead of the
algorithm returning an explicit value. As we are dealing with maximising tree
traversals, bounds are always compared using a \emph{greater than} $(>)$ function defined on the
\emph{Bound} type.

Parallelism may be introduced introduced by searching the set of candidates
\emph{speculatively} in parallel, as illustrated in~\cref{sec:skeletons}.
Parallel search branches allow early updates of the incumbent via a synchronised
version of the solution read/write interface.

\begin{algorithm}
  \DontPrintSemicolon
  $\FuncSty{expandSearch}$ (space, node)\;
  \Begin{
    candidates = orderedGenerator(space, node) \;
    \If{null(candidates)}{
      return \tcp*{Backtrack}
    }
    \tcp{Parallelism may be introduced here}
    \For{c in candidates} {
      bestBound = currentBound() \tcp*{built-in function}
      localBest = pruningHeuristic(space, node)
      \If{localBest \(>\) bestBound} {
        \If{bound(node) \(>\) bestBound)} {
            updateBest(solution(node), bound(node)) \tcp*{built-in function}
        }
        expandSearch(space, c) \;
      }
    return \tcp*{Backtrack}
    }
  }
\caption{General algorithm for branch and bound search using the GBB API (\cref{lst:BaBAPI})}
\label{alg:parallelCoordination}
\end{algorithm}

\subsection{Implementing the GBB API}
\label{sec:GBB-optimisations}

Although GBB  can encode general branch and bound searches,
various modifications improve both sequential and parallel efficiency.

Generally the search space is immutable and fixed at the start of the
search. In a distributed environment we can avoid copying the search space each time a task is stolen by storing a read only copy of the search space on each host.
It is also possible to remove the space argument
from the API functions and add accessor functions in the same manner as bound
access. The implementations used in~\cref{sec:evaluation} do pass the
space as a parameter.

For some applications, such as Maximum Clique, if the local bound fails to
unseat the incumbent then all other candidate nodes \emph{to-the-right}
(assuming an ordered generator) will also fail the pruning predicate. An
implementation can take advantage of this fact and break the candidate checking
loop for an early backtrack. This optimisation is key in avoiding wasteful
search. In the skeleton implementations used in~\cref{sec:evaluation} we allow
this behaviour to be toggled via a runtime flag.

Finally, an implementation can exploit lazy evaluation within the node type to
avoid redundant computation. Taking Maximum Clique as an example we can delay
the computation of the set of candidates vertices until after the pruning
heuristic has been checked (as this only depends on having the bound and
colour). Similarly if we use the to-the-right pruning optimisation, described
above, we want to avoid paying the cost of generating the nodes which end up
being pruned.

\section{Parallel Skeletons for Branch and Bound Search}
\label{sec:skeletons}

Algorithmic skeletons are higher order functions that abstract over
common patterns of coordination and are parameterised with specific
computations~\cite{cole_skeletons:1991}. For example, a parallel map
function will apply a sequential function to every element of a
collection in parallel. Skeletons are polymorphic, so the collection
may contain elements of any type, and the function type must match the
element type. The programmer's task is greatly simplified as they do
not need to specify the coordination behaviour required.  The skeleton
model has been very influential, appearing in parallel standards such
as MPI and OpenMP~\cite{MPI,OpenMP}, and distributed skeletons such as
Google's MapReduce~\cite{MapReduceCACM} are core elements of cloud
computing.

Here the focus is on designing skeletons for maximising branch and
bound search on distributed memory architectures. These architectures
use multiple cooperating processes with distinct memory
spaces. The processes may be spread across multiple hosts.

Although it is possible to implement skeletons using a variety of parallelism
models, we adopt a task parallel model here. The task parallel model is based
around breaking down a problem into multiple units of computation (\emph{tasks})
that work together to solve a particular problem. In a distributed setting,
tasks (and their results) may be shared between processes. For search trees,
parallel tasks generally take the form of sub-trees to be searched.

Two skeleton designs are given in this section. The first skeleton,
\textbf{Unordered}, makes no guarantees on the search ordering and so may give
the anomalous behaviours and the unpredictable parallel performance outlined in
~\cref{sec:searchAnomalies}. The second skeleton, \textbf{Ordered}, enforces a
strict search ordering and hence avoids search anomalies and gives predictable
performance. The unordered skeleton is used as an example of the pitfalls
of using a standard random work stealing approach and provides a baseline
comparison for evaluating the performance of the Ordered skeleton
(\cref{sec:evaluation}).

We start by considering the key design choices for constructing a branch and
bound skeleton. Using these we show how the Unordered skeleton can be
constructed, and then show the modifications required to transform the
Unordered into the Ordered skeleton. \cref{sec:skelcomparison} summarises the
design choices and limitations of the design choices are summarised
in~\cref{sec:skelLimitations}.

\subsection{Design Choices}

Three main questions drive the design of branch and bound search skeletons:

\begin{enumerate}
  \item How is work generated?
  \item How is work distributed and scheduled?
  \item How are the bounds propagated?
\end{enumerate}

The first two choices focus on task parallel aspects of the design and are
common design features for algorithmic skeletons. Bound propagation is a specific
issue for branch and bound search and takes the form of a general coordination
issue rather than being tied to the task parallel model.

To achieve performance in the task parallel model, tasks should be
oversubscribed, that is there should be more tasks than cores, while avoiding
low task granularity where communication and synchronisation overheads may
outweigh the benefits of the parallel computation. To achieve these
characteristics in the skeleton designs a simple approach for work generation is
used: generate parallel tasks from the root of the tree until a given depth
threshold is reached. This method exploits the heuristic that tasks near the top
of the tree are usually of coarse granularity than those nearer the leaves,
i.e.\ they have more of the search space to consider. This threshold approach is
commonly used in divide-and-conquer parallelism and allows a large number of
tasks to be generated while avoiding low granularity tasks. The argument that
tasks near the top of the tree have coarse granularity does not necessarily hold
true for all branch and bound searches as variant candidate sets and pruning can
truncate some searches initiated near the root of the tree: hence task
granularity may be highly irregular.

\subsection{Unordered Skeleton}

The type signature of the Unordered skeleton is:

\begin{minipage}{\textwidth}
\begin{haskellcodenofloat}
search :: Int  -- Depth to spawn to
       -> Node Sol Bnd Candidates -- Root node
       -> (Space -> Node Sol Bnd Candidates -> [Node Sol Bnd Candidates]) -- orderedGenerator
       -> (Space -> Node Sol Bnd Candidates -> Bool) -- pruningHeuristic
       -> Par Solution
\end{haskellcodenofloat}
\end{minipage}

In the skeleton search tasks recursively generate work, i.e.\ new search
tasks.  If the depth of a search task does not exceed the threshold it
generates new tasks on the host, otherwise the task searches the
subtree sequentially.

Work distribution takes the form of random work stealing with
exponential back-off~\cite{blumofe.joerg.ea:1996} and happens at two
levels.  Intranode steals occur between two workers in the same
process, the next sub-tree is stolen from the workqueue of the local
process.  Only if the worker fails to find local work does an
internode steal occur, targeting some random other process.  Only one
internode steal per process is performed at a time.  New tasks, either
created by local workers or stolen from remote processes, are added to
the local workqueue and are scheduled in last-in-first-out order.

The current incumbent, i.e. best solution, is held on every host, and managed by
a distinguished master process. Bound propagation proceeds in two stages.
Firstly when a search task discovers a new Solution it sends both the solution
and bound to the master and, if no better solution has yet been found, they
replace the incumbent. Secondly the master broadcasts the new bound to all other
processes, that update their local incumbent unless they have located a better
solution. This is a form of eventual
consistency~\cite{vogels_eventuallyConsistent:2009} on the incumbent. Using this
approach, opposed to fully peer to peer, the new solution is sent to the master
once and only bounds are broadcast. While broadcast is bandwidth intensive,
broadcasting new bounds provides fast knowledge transfer between search tasks.
Moreover experience shows that often a \emph{good}, although not necessarily
optimal, bound is found early in the search making bound updates rare. In many
applications the bounds are range-limited, e.g.\ a Maximum Clique cannot be
larger than the number of vertices in the graph.

\subsection{Ordered Skeleton}

The type signature of the Ordered skeleton is as follows.

\begin{minipage}{\textwidth}
\begin{haskellcodenofloat}
search :: Bool -- Diversify search
       -> Int  -- Depth to spawn to
       -> Node Sol Bnd Candidates -- Root node
       -> (Space -> Node Sol Bnd Candidates -> [Node Sol Bnd Candidates]) -- orderedGenerator
       -> (Space -> Node Sol Bnd Candidates -> Bool) -- pruningHeuristic
       -> Par Solution
\end{haskellcodenofloat}
\end{minipage}

The additional first parameter enables discrepancy search ordering
(\cref{sec:eval-maxclique}) to be toggled; an alternative formulation would be
to pass an ordering function in explicitly. The skeleton adapts the Unordered skeleton to avoid search anomalies
(\cref{sec:searchAnomalies}) and give predictable performance properties as
shown in~\cref{sec:introduction}.

The {\propA} property guarantees that parallel runtimes do not exceed the
sequential runtime. To maintain this property we enforce that at least one
worker executes tasks in the exact same order as the sequential search. The
other workers speculatively execute other search tasks and may improve the bound
earlier than in the fully sequential case, as illustrated in
\cref{fig:ExampleReduction}. Discovering a better incumbent early enables the
sequential thread to prune more aggressively and hence explore less of the tree
than the entirely sequential search would, providing speedups. While there is no
guarantee that the speculative workers will improve the bound, the property will
still be maintained by the sequential worker.

Requiring a sequential worker is a departure from the fully random
work stealing model. Instead of all workers performing random steals,
the task scheduling decisions are enforced for the sequential worker.
Our system achieves sequential ordering by duplicating the task
information. One set is stealable by any worker, and the other is
restricted to the sequential worker. There is a chance that work will
be duplicated as some worker may simultaneously attempt to start the same
task as the sequential worker. To avoid duplicating work, we use a
basic locking mechanism where workers first check whether a task
has already started execution before starting the task themselves.

With random scheduling adding a worker may disrupt a good parallel
search order (\cref{sec:searchAnomalies}), so to guarantee the
\textbf{non-increasing runtimes} property we need to preserve the
parallel search order, just as the sequential worker preserves the
sequential search order. Preserving the parallel search order means
that if with $n$ workers we locate an incumbent by time $t_{pn}$, then
with $n+1$ workers we locate the same incumbent, or a better
incumbent, at approximately $t_{pn}$. The approximation is required
as, in a distributed setting, $t_{pn}$ may vary slightly due to the
speed of bound propagation.

It transpires that preserving the parallel search order is also
sufficient to guarantee the \textbf{repeatability} property as all
parallel executions follow very similar search orders. The parallel search order must be globally visible for it to be
preserved, and we can no longer permit random work stealing. Instead
all tasks are generated on the master host and maintained in a central \emph{priority} queue.
In our skeleton implementation we use depth-bounded work
generate to statically construct a fixed set of tasks, with set
priorities, before starting the search. Alternative
work generation approaches, for example  dynamic generation, are possible provided all  tasks
are generated on the master host.

The parallel search order may have dramatic effects on search
performance~\cite{ChuSS09,MoisanQG14,mccreesh.prosser_shapeOfTheSearchTree:2015}.
In our skeletons \emph{any} fixed ordering will maintain the properties,
although it may not guarantee good performance. The GBB API
in~\cref{sec:genericBBSearch} relies on the user choosing an ordering of nodes
in the \emph{orderedGenerator} function. This ordering is generally, but not
necessarily, based on some domain specific heuristic. One simple scheduling
decision, and our default, is to assign priorities from left-most to right-most
task in the tree. The skeleton may use any priority order rather than the
default left-to-right order, for example the depth-bounded discrepancy (DDS)
order~\cite{walsh_DDS:1997}. This discrepancy ordering is used when
evaluating the Maximum Clique benchmark (\cref{sec:eval-maxclique}).

By augmenting the Unordered skeleton with a single worker that follows the
sequential ordering and a global priority ordering on tasks we arrive at the
Ordered skeleton that provides reliable performance guarantees while still
enabling parallelism.

\subsection{Skeleton Comparison}
\label{sec:skelcomparison}

\cref{tab:skeletonFeatureComparison} compares the key design features
of the two skeletons. A key difference is where tasks are generated
and stored.  The Unordered skeleton adopts a dynamic approach at the
cost of not giving the same performance guarantees as the Ordered
skeleton due to a lack of global ordering.  Many other skeleton
designs are possible. An advantage of the skeleton approach that
exploits a general API is that parallel coordination alternatives may
be explored and evaluated without refactoring the application code.

\begin{table}[h]
  \centering
  \begin{tabular}{@{}l l l@{}}
    \toprule
                                & \textbf{Unordered}                     & \textbf{Ordered}                     \\
    \midrule
    \textbf{Work Generation}    & Dynamically to depth \(d\) on any host & Statically to depth \(d\) on master  \\
    \textbf{Work Distribution}  & Random work stealing all processes & Work stealing master process only  \\
    \textbf{Bounds Propagation} & Broadcast                              & Broadcast                            \\
    \textbf{Sequential Worker}  & False                                  & True                                 \\
    \bottomrule
  \end{tabular}
  \caption{Skeleton comparison}
  \label{tab:skeletonFeatureComparison}
\end{table}

\subsection{Limitations}
\label{sec:skelLimitations}

For most design choices we have selected a simple alternative. More
elaborate alternatives might well deliver better performance. Here we
discuss some of the limitations imposed by the simple alternatives
selected.

One key limitation of both skeleton designs is the use of depth bounded work
generation techniques. While this technique is a well known optimisation for
divide and conquer applications, the need to manually tune the depth threshold
reduces the skeleton portability as the number of tasks required to populate a
system is proportional to the system size. Given the irregular structure of a
branch and bound computation it is often difficult to know ahead of time how
many tasks will need to be generated to avoid starvation and fully exploit the
resources available. In practise we have not found this to be an issue, as for
many problem instances such as the three benchmarks used in the skeleton
evaluation (\cref{sec:applications}), even generating work to a depth of 1 can
give thousands of tasks. However, for some instances, to achieve best
performance one may need to split work at much lower
levels~\cite{mccreesh.prosser_shapeOfTheSearchTree:2015}. An alternative would
be to use dynamic work generation techniques where the parallel coordination
layer manages load in the system ~\cite{abukhzam_scalableParallelBacktracking}.
Dynamic work generation can cause difficulty for maintaining a global task
ordering in a distributed environment such as in the case of the Ordered
skeleton.

A consequence of static work generation in the Ordered skeleton is
that the runtime for the single worker case can be larger than that of a fully
sequential search implementation. With static work generation, work is generated
from nodes at a depth \(d\) ahead of time and the parent nodes are no longer
considered (as they are already searched). This leads to the creation of
additional tasks that a sequential implementation may never create due to
pruning at the higher levels. The management and searching of these additional
tasks causes the discrepancy between the single worker Ordered skeleton and
purely sequential search. While this does not effect the properties, as we
phrase property 1 in terms of a single worker, it would if a
purely sequential implementation in property 1 is considered. The effects of this limitation
could be mitigated by treating all nodes above the depth threshold as tasks and
allowing cancellation of parent/child tasks. Such an approach complicates the
task coordination greatly as tasks require knowledge of both their parent and
child task states.

The Ordered skeleton requires additional memory and processing time on
the master host to maintain the global task list and respond promptly
to work stealing requests.  In practise we have not found this to be a
significant issue as most tasks near to top of the search tree are
long running and the steals occur at irregular intervals.  On large
distributed systems, and for some searches, it is possible that a
single master might prove to be a scalability bottleneck.

\subsection{Implementation}
\label{sec:hdphImplementation}

The Ordered and Unordered skeletons are implemented in \emph{Haskell distributed
  parallel Haskell} (HdpH) embedded Domain Specific Language
(DSL)~\cite{maier.stewart.ea:2014}. HdpH has been modified to use a priority
queue based scheduler to enable the strict ordering on task execution. While
HdpH cannot match the performance of the state of the art branch and bound
search implementations it is useful for evaluating the skeletons for the
following reasons.

\begin{enumerate}
\item HdpH supports the higher order functions, a commonly used approach for constructing skeletons.
\item The HdpH is small and easy to modify,  allowing ideas to
  be rapidly prototyped. For example we experimented with priority-based work stealing.
\item The properties of the Ordered skeleton depend on relative runtime values, i.e.\ absolute runtime is not the priority.
\end{enumerate}

Although our skeletons have been implemented in a functional language they may
be implemented in any system with the following features: task parallelism;
work stealing (random/single-source); locking; priority based work-queues/task
ordering. Distributed memory skeleton implementations will also require
distribution mechanisms and distributed locking.

\subsection{Maximum Clique Representation}

To end this section we show, using the functions and types defined in
\cref{lst:maxclique-impl} how the search skeletons are used within an
application. Here we show how the skeleton is called for the Maximum Clique
benchmark (\cref{sec:motivatingExample}):

\begin{minipage}{\textwidth}
  \begin{minipage}{.45\textwidth}
    \begin{haskellcodenofloat}
    Unordered.search
      spawnDepth
      (Node ([], 0), 0. allVertices)
      orderedGenerator
      pruningHeuristic
    \end{haskellcodenofloat}
  \end{minipage}
  \begin{minipage}{.45\textwidth}
    \begin{haskellcodenofloat}

    Ordered.search
      True -- Use discrepancy search
      spawnDepth
      (Node ([], 0), 0. allVertices)
      orderedGenerator
      pruningHeuristic
    \end{haskellcodenofloat}
  \end{minipage}
\end{minipage}

\subsection{Other Branch and Bound Skeletons}

While algorithmic skeletons are widely used in a range of areas from processing
large data sets~\cite{MapReduceCACM} to multicore programming~\cite{intelTBB}
there has been little work on branch and bound skeletons. Two notable exceptions
are MALLBA~\cite{alba.almeida.ea_MALLBAALibraryOfSkeletons:2002} and
Muesli~\cite{poldner.kucher_AlgorithmicSkeletonsForBranchAndBound:2006} that
both provide distributed branch and bound implementations. Both frameworks are
written in C++.
Muesli uses a similar \emph{higher-order function} approach to ourselves while MALLBA is designed around using \emph{classes} and \emph{polymorphism} to override solver behaviour.
In Muesli it is possible to choose between a
centralised workpool approach, similar to the Ordered skeleton but using
work-pushing rather than work stealing, or a distributed method. Unfortunately
the centralised workpool model does not scale well compared with our approach
(\cref{sec:evaluation}). MALLBA similarly uses a single, centralised, workqueue
for its branch and bound implementation. The real strength of the MALLBA
framework is in the ability to encode multiple exact and inexact combinatorial
skeletons as opposed to just branch and bound.

The Muesli authors further highlight the need for reproducible runtimes and note
``the parallel algorithm behaves non-deterministically in the way the
search-space tree is explored. In order to get reliable results, we have
repeated each run 100 times and computed the average
runtimes''~\cite{poldner.kucher_AlgorithmicSkeletonsForBranchAndBound:2006}. By
adopting the strictly ordered approach in this paper we avoid the need for large
numbers of repeated measurements to account for non-deterministic search
ordering.

\section{Model, API and Skeleton Generality}
\label{sec:applications}

To show that the BBM model and GBB API are generic, and to provide additional
evidence that the Ordered skeleton preserves the parallel search properties
(\cref{sec:evaluation}) we consider two additional search benchmarks: 0/1 Knapsack, a binary assignment problem, and travelling salesperson, a
permutation problem.

\subsection{0/1 Knapsack}
\label{sec:applications-knapsack}

Knapsack packing is a classic optimisation problem. Given a container of some
finite size and a set of items, each with some size and value, which items
should be added to the container in order to maximise its value? Knapsack problems
have important applications such as bin-packing and industrial decision making
processes~\cite{salkin.dekluyver_knapsackSurvey:1975}. There are many variants
of the knapsack problem~\cite{martello.toth_knapsack:1990}, typically changing
the constraints on item choice. For example we might allow an item to be chosen
multiple times, or fractional parts of items to be selected. We consider the 0/1
knapsack problem where an item may only be selected once and fractional items
are not allowed.

At each step a bound may be calculated using a linear relaxation of the
problem~\cite{martousek_ILP} where, instead of solving for $i \in \{0,1\}$
we instead solve fractional knapsack problem where $i \in [0,1]$. As
the greedy fractional approach is optimal and provides an upper bound
on the maximum potential value. Although it is possible to compute an
upper bound on the entire computation by considering the choices at
the top level~\cite{martello.toth_upperBoundKnapsack}, we do not this
here.  The primary benefit of this method is to terminate the search
early when a maximal solution is found.

A formalisation of the 0/1 Knapsack problem in BBM and the corresponding
GBB implementation are given in \cref{sec:formalKnapsack} and
\cref{sec:GBBknapsack} respectively.

\subsection{Travelling Salesperson Problem}
\label{sec:applications-tsp}

Travelling salesperson (TSP) is another classic optimisation problem.
Given a set of cities to visit and the distance between each city find the
shortest tour where each city is visited once and the salesperson returns to the
starting city. We consider only \emph{symmetric} instances where the distance
between two cities is the same travelling in both directions.

A formalisation of TSP in BBM and the corresponding GBB implementation
are given in \cref{sec:formalTSP} and \cref{sec:GBBTSP} respectively.

\section{Parallel Search Evaluation}
\label{sec:evaluation}

This section evaluates the parallel performance of the Ordered and Unordered
generic skeletons. It starts by outlining the benchmark instances
(\cref{sec:eval-instances}) and experimental platform (\cref{sec:platform}). We
establish a baseline for the overheads of the generic skeletons by comparing
them with a state of the art C++ implementation (\cref{sec:seq-eval}) of Maximum
Clique. Finally we investigate the extent that the Ordered skeleton preserves
the runtime and repeatability properties (\cref{sec:properties}) for the three
benchmarks.

The datasets supporting this evaluation are available from an open access archive~\cite{archibald-JPDC-dataset}.

\subsection{Benchmark Instances and Configuration}
\label{sec:eval-instances}

This section specifies how the benchmarks are configured and the
instances used.  We aim for test instances with a runtime of less than
an hour while avoiding short sequential runtimes that don't benefit
from parallelism. These instances ensure we a) have enough parallelism
and b) can perform repeated measurements while keeping computation
times manageable.

\subsubsection{Maximum Clique}
\label{sec:eval-maxclique}

The Maximum Clique implementation (\cref{sec:motivatingExample}) measured uses
the bit set encoded algorithm of San Segundo et al:
BBMC~\cite{segundo.losada.ea_BBMC:2011, segundo.matia.ea_improvedBBMC:2011}.
This algorithm makes use of bit-parallel operations to improve performance in
the greedy colouring step (\textit{orderedGenerator} in the GBB API), and ours
is the first known \textit{distributed parallel} implementation of BBMC. We do
not use the additional recolouring
algorithm~\cite{segundo.matia.ea_improvedBBMC:2011}. Maximum Clique is one
example where prunes can propagate to-the-right (\cref{sec:GBB-optimisations})
and we make use of this in the implementation. The instances are given in
\cref{tab:mcinstances} and come from the second DIMACS implementation
challenge~\cite{secondDimacsChallenge}.

\begin{table}[h]
  \centering
  \begin{tabular}{llll}
    \toprule
    brock400\_1 & brock800\_1 & MANN\_a45    & sanr200\_0.9 \\
    brock400\_2 & brock800\_2 & p\_hat1000-2 & sanr400\_0.7 \\
    brock400\_3 & brock800\_3 & p\_hat500-3  & \\
    brock400\_4 & brock800\_4 & p\_hat700-3  & \\
    \bottomrule
  \end{tabular}
  \caption{Maximum Clique instances}
  \label{tab:mcinstances}
\end{table}

\begin{figure}
  \centering
    \begin{tikzpicture}[level 1/.style={sibling distance=7em}, level 2/.style={sibling distance=2.5em}]
      \node[choice=4] (lvl1){\nodepart{two}\nodepart{three}\nodepart{four}\nodepart{five}\nodepart{six}}
      child {
        node[choice=3] (1_1) {\nodepart{two}\nodepart{three}}
        edge from parent[draw=none]
        child {
          node[choice=2] (2_1) {0\nodepart{two}4\nodepart{three}}
          edge from parent[draw=none]
        }
        child {
          node[choice=1] (2_2) {3\nodepart{two}}
          edge from parent[draw=none]
        }
        child {
          node[choice=2] (2_3) {4\nodepart{two}7\nodepart{three}}
          edge from parent[draw=none]
        }
      }
      child {
        node[choice=2] (1_2) {\nodepart{two}\nodepart{three}}
        edge from parent[draw=none]
        child {
          node[choice=2] (2_4) {1\nodepart{two}5\nodepart{three}}
          edge from parent[draw=none]
        }
        child {
          node[choice=1] (2_5) {3\nodepart{two}}
          edge from parent[draw=none]
        }
      }
      child {
        node[choice=1] (1_3) {}
        edge from parent[draw=none]
        child {
          node[choice=1] (2_6) {2\nodepart{two}}
          edge from parent[draw=none]
        }
      }
      child {
        node[choice=2] (1_4) {\nodepart{two}\nodepart{three}}
        edge from parent[draw=none]
        child {
          node[choice=1] (2_7) {3\nodepart{two}}
          edge from parent[draw=none]
        }
        child {
          node[choice=2] (2_8) {5\nodepart{two}9\nodepart{three}}
          edge from parent[draw=none]
        }
      };

      \node[moreTree] at (2_1.one south) {};
      \node[moreTree] at (2_1.two south) {};
      \node[moreTree] at (2_2.south) {};
      \node[moreTree] at (2_3.one south) {};
      \node[moreTree] at (2_3.two south) {};
      \node[moreTree] at (2_4.one south) {};
      \node[moreTree] at (2_4.two south) {};
      \node[moreTree] at (2_5.south) {};

      \node[moreTree] at (2_6.south) {};
      \node[moreTree] at (2_7.south) {};
      \node[moreTree] at (2_8.one south) {};
      \node[moreTree] at (2_8.two south) {};

      \draw[] (lvl1.one south) -- (1_1.north);
      \draw[] (lvl1.two south) -- (1_2.north);
      \draw[] (lvl1.three south) -- (1_3.north);
      \draw[] (lvl1.four south) -- (1_4.north);

      \draw[] (1_1.one south) -- (2_1.north);
      \draw[] (1_1.two south) -- (2_2.north);
      \draw[] (1_1.three south) -- (2_3.north);

      \draw[] (1_2.one south) -- (2_4.north);
      \draw[] (1_2.two south) -- (2_5.north);

      \draw[] (1_3.one south) -- (2_6.north);

      \draw[] (1_4.one south) -- (2_7.north);
      \draw[] (1_4.two south) -- (2_8.north);

  \end{tikzpicture}
  \caption{Discrepancy search priorities - lower is higher priority}
  \label{fig:DDS_priorities}
\end {figure}
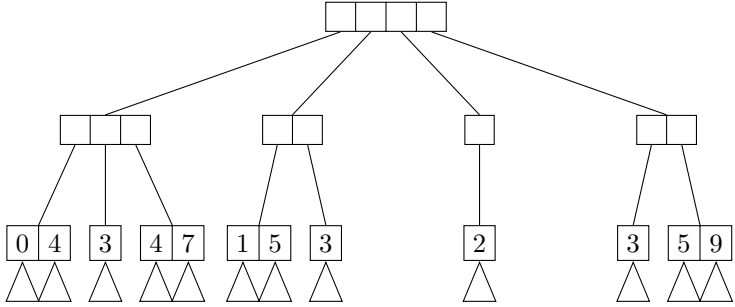

For many applications, search heuristics are weak and tend to perform badly near
the root of the search tree~\cite{HarveyG95}. To overcome this limitation, the
Maximum Clique example makes use of a non left-to-right search ordering in order
to make the search as diverse as possible. The new order is based on
depth-bounded discrepancy search~\cite{walsh_DDS:1997} with the algorithm
extended to work on n-ary trees by counting the nth child as n discrepancies. An
example of the discrepancy search ordering is shown
in~\cref{fig:DDS_priorities}\footnote{Different discrepancy orderings can exist
  depending on how discrepancies are counted and which biases are applied.}.
This further shows the generality of the skeleton to maintain the properties
even when custom search orderings are used.

\subsubsection{The 0/1 Knapsack Problem}
\label{sec:eval-knapsack}

The 0/1 Knapsack implementation (\cref{sec:applications-knapsack}) uses ascending
profit density ordering as the search heuristic and a greedy fractional knapsack
implementation for calculating the lower bound. As with Maximum Clique we take
advantage of the prune to-the-right optimisation. The bound is uninitialised at
the start the search. This simple implementation does not match the performance
of state of the art solvers.

Although the knapsack problem is NP-hard, many knapsack instances are
easily solved on modern hardware. Methods exist for generating \textit{hard}
knapsack instances~\cite{pisinger_whereAreTheHardKnapsackProblems:2015}. We make
use of the subset of the pre-generated hard instances~\cite{pisingerInstances}
shown in~\cref{tab:knapsackInstances}.

\begin{table}
  \centering
  \begin{tabular}{llr}
    \toprule
    Instance Name (Pisinger) & Type & Number of Items \\
    \midrule
    knapPI\_11\_100\_1000\_37 & Uncorrelated span(2,10) & 100 \\
    knapPI\_11\_50\_1000\_40 & Uncorrelated span(2,10) & 50 \\
    knapPI\_12\_50\_1000\_23 & Weakly Correlated span(2,10) & 50 \\
    knapPI\_12\_50\_1000\_34 & Weakly Correlated span(2,10) & 50 \\
    knapPI\_13\_50\_1000\_10 & Strongly Correlated span(2,10) & 50 \\
    knapPI\_13\_50\_1000\_32 & Strongly Correlated span(2,10) & 50 \\
    knapPI\_14\_100\_1000\_88 & Multiple Strongly Correlated & 100 \\
    knapPI\_14\_50\_1000\_64 & Multiple Strongly Correlated & 50 \\
    knapPI\_15\_500\_1000\_47 & Profit Ceiling & 500 \\
    knapPI\_15\_50\_1000\_20 & Profit Ceiling & 50 \\
    knapPI\_16\_50\_1000\_62 & Circle & 100 \\
    knapPI\_16\_50\_1000\_21 & Circle & 50 \\
    \bottomrule
  \end{tabular}
  \caption{0/1 Knapsack instances}
  \label{tab:knapsackInstances}
\end{table}

\subsubsection{Travelling Salesperson}
\label{sec:eval-tsp}

The final application is the travelling salesperson problem
(\cref{sec:applications-tsp}). A simple implementation is used that assumes no
ordering on the candidate cities and uses Prim's minimum spanning tree
algorithm~\cite{prim} to construct a lower bound. The initial bound comes from
the result of a greedy nearest neighbour search.

Like the knapsack application, this is a proof of concept implementation, based
on simple branching and pruning functions, and does not perform as well as
current state of the art solvers which go beyond simple branch and bound search.

Problem instances are drawn from two separate locations: the TSPLib
instances~\cite{TSPLib} and random instances from the DIMACS TSP challenge
instance generator~\cite{DIMACS-TSP}. A list of benchmarks is given
in~\cref{tab:tspInstances}.

\begin{table}
  \centering
  \begin{tabular}{llll}
    \toprule
    Name & Type & Cities & Random Seed \\
    \midrule
    burma14 & TSPLib & 14 & \\
    ulysses16 & TSPLib & 16 & \\
    ulysses22 & TSPLib & 22 & \\
    rand\_1 & DIMACS Challenge & 34 & 22137 \\
    rand\_2 & DIMACS Challenge & 35 & 52156 \\
    rand\_3 & DIMACS Challenge & 35 & 52156 \\
    rand\_3 & DIMACS Challenge & 36 & 62563 \\
    rand\_4 & DIMACS Challenge & 37 & 6160  \\
    rand\_5 & DIMACS Challenge & 38 & 37183 \\
    rand\_6 & DIMACS Challenge & 39 & 50212 \\
    \bottomrule
  \end{tabular}
  \caption{TSP instances}
  \label{tab:tspInstances}
\end{table}

\subsection{Measurement Platform and Protocols}
\label{sec:platform}

The evaluation is performed on a Beowulf cluster consisting of 17 hosts each
with dual 8-core Intel Xeon E5-2640v2 CPUs (2Ghz), 64GB of RAM and running
Ubuntu 14.04.3 LTS. Exclusive access to the machines is used and we ensure there
is always at least one physical core per thread. Threads are assigned to cores
using the default mechanisms of the GHC runtime system.

The skeleton library and applications are written in Haskell using the
HdpH distributed-memory parallelism framework as outlined
in~\cref{sec:hdphImplementation}. Specifically we use the GHC 8.0.2
Haskell compiler and dependencies are pulled from the stackage lts-7.9
repository or fixed commits on github\footnote{See stack.yaml at
  \url{http://dx.doi.org/10.5281/zenodo.254088} for details of the
  dependencies}.  The complete source code for the experiments is
available at: \url{http://dx.doi.org/10.5281/zenodo.254088}.

In all experiments, each HdpH node (runtime) is assigned \(n\) threads and
manages \(n-1\) workers that execute the search. The additional thread is used
for handling messages from other processes and garbage collection and does not
search. The additional thread minimises the performance impact of overheads like
communication and garbage collection. Measurements are taken with 1, 2, 4, 8,
32, 64, 128 and 200 workers.

Unless otherwise specified, all results are based on the mean of ten runs.
The \texttt{spawnDepth} is always set to one, causing child tasks to be spawned for each top level task.
This \texttt{spawnDepth} setting provided good performance for most instances, however it may not be optimal for each individual instance.

\subsection{Comparison with a Class-leading C++ Implementation}
\label{sec:seq-eval}

To establish a performance baseline for the generic Haskell skeletons we compare
the sequential (single worker) performance of the skeletons with a state of the
art C++ implementation of the Maximum Clique
benchmark~\cite{mccreesh.prosser_shapeOfTheSearchTree:2015}. Only instances with
a (skeleton) sequential runtime of less than one hour are considered.

The C++ results were gathered on a newer system featuring a dual Intel Xeon
E5-2697A v4, 512 GBytes of memory, Ubuntu 16.04 and were compiled using g++ 5.4.
A single sequential sample is used for comparison.

\Cref{tab:sequentialComparison} compares the C++ implementation to the Ordered
skeleton. To keep the skeleton execution as close to a fully sequential
implementation as possible, work is generated only at the top level and is
searched in decreasing degree order. As there is no communication, the HdpH node
is assigned a single thread and a single worker.

\begin{table}
  \centering
  \begin{tabular}{lrrr}
    \toprule
    \textbf{Instance} & \textbf{C++ (s)} & \textbf{Ordered Skeleton (s)} & \(\frac{\textbf{Ordered Skeleton}}{\textbf{C++}}\)\\
    \midrule
    brock400\_1        & 184.4 &  987.7 & 5.36 \\
    brock400\_2        & 133.7 &  725.8 & 5.43 \\
    brock400\_3        & 106.1 &  577.7 & 5.44 \\
    brock400\_4        & 51.6  &  275.5 & 5.34 \\
    MANN\_a45          & 123.2 &  238.2 & \textbf{1.93} \\
    p\_hat1000-2       & 95.0  &  421.8 & 4.44 \\
    p\_hat500-3        & 70.9  &  368.1 & 5.19 \\
    sanr200\_0.9       & 14.3  &  88.1  & \textbf{6.16} \\
    sanr400\_0.7       & 48.3  &  274.7 & 5.69 \\
    \bottomrule
  \end{tabular}
  \caption{Sequential Runtimes of a Class-leading C++ search and the Generic Haskell Ordered Skeleton}
  \label{tab:sequentialComparison}
\end{table}

As expected, the Ordered skeleton is between a factor of 1.9 and 6.2
slower than the hand crafted C++ search.  A primary contributor to the slowdown
is Haskell execution time: with the slowdown widely accepted to be a
factor of between 2 to 10, but often lower for symbolic computations
like these. The slowdown is due to Haskell's aggressive use of
immutable heap structures, garbage collection and lazy evaluation
model. The generality of the skeletons means that they use
computationally expensive techniques like higher-order functions and
polymorphism. Finally, our skeleton implementations have not been
extensively hand optimised, as the C++ implementation has.

The remainder of the evaluation uses speedup relative to the one worker Haskell implementation.
We argue that the underlying performance in the sequential (one worker) is sufficiently good for the results to be credible.

\subsection{{\propA} \& {\propB}}
\label{sec:PropAPropB}
As {\propA} and {\propB} are both runtime properties we evaluate
them together. We investigate the relative speedup, or strong scaling, of the
Ordered and Unordered skeletons using between 1 and 200 workers for
each benchmark. If {\propA} holds then the speedup will be greater than or equal to 1, and
if {\propB} holds the curves should be non-decreasing. {\propB} is
still maintained even when a speedup curve becomes flat: we simply
don't benefit from additional workers.

\begin{figure}[h]
  \centering
  \includegraphics[width=0.55\textwidth]{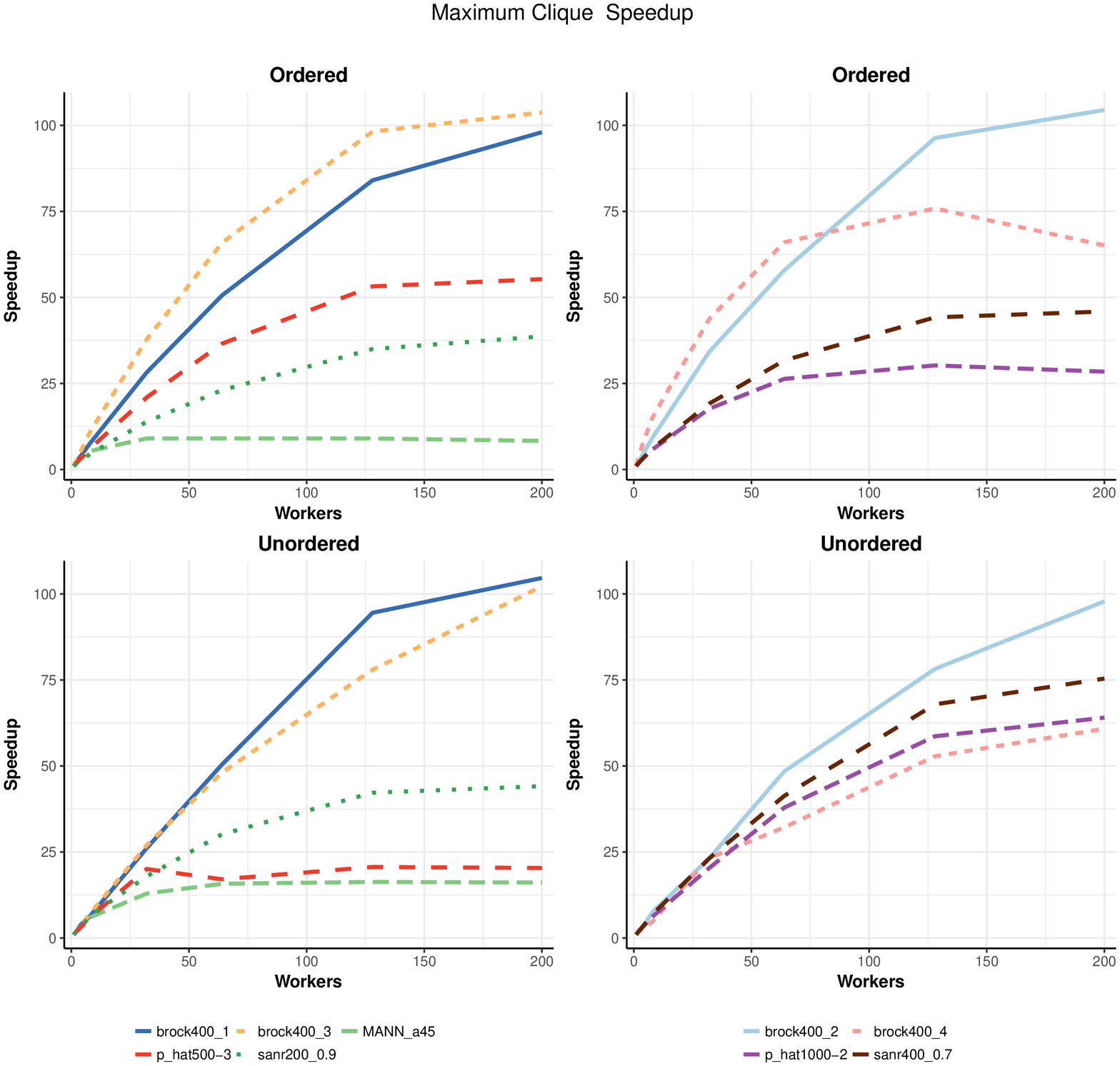}
  \vspace{-0.3cm}
  \caption{Maximum Clique Speedups: Ordered Skeleton Maintains {\propA} and {\propB} Properties}
  \label{fig:MC_Speedup}
\end{figure}

\Cref{fig:MC_Speedup} shows the speedup curves for the Maximum Clique Ordered
and Unordered skeletons. Scaling curves are not given for the brock800 series
and the p\_hat700-3 instances as instances with a one worker baseline of greater
than one hour are not considered.

For all Maximum Clique instances both skeletons preserve {\propA}, i.e.\ no
configuration has greater runtime than the single worker case. The skeletons
achieve good parallel speedups for symbolic computations, delivering a maximum
parallel efficiency of around 50\%.

The Ordered Skeleton maintains {\propB} for most instances, exceptions being brock400\_4, p\_hat100-2 and MANN\_a45, shown by non-decreasing speedup curves.
brock400\_4 appears to have the largest slow down between 128 to 200 workers, however the runtime at these scales is tiny (4.5s), and we attribute the slowdown to a combination of (small) parallelism overheads and variability.
While the final runtimes for p\_hat1000-2 and MANN\_a45, once parallelism stops being effective, are larger (15s and 27s respectively) the mean runtimes for 64, 128 and 200 workers are never more than 2.5 seconds apart and this we again attribute to parallelism overheads rather than search ordering issues.
Even for instances such as MANN\_a45, where there is limited performance benefit for adding additional workers due to a large maximum clique causing increased amounts of pruning, using additional cores never increases runtime significantly.

\begin{figure}[h]
  \centering
  \begin{subfigure}[b]{0.3\textwidth}
    \includegraphics[width=\textwidth]{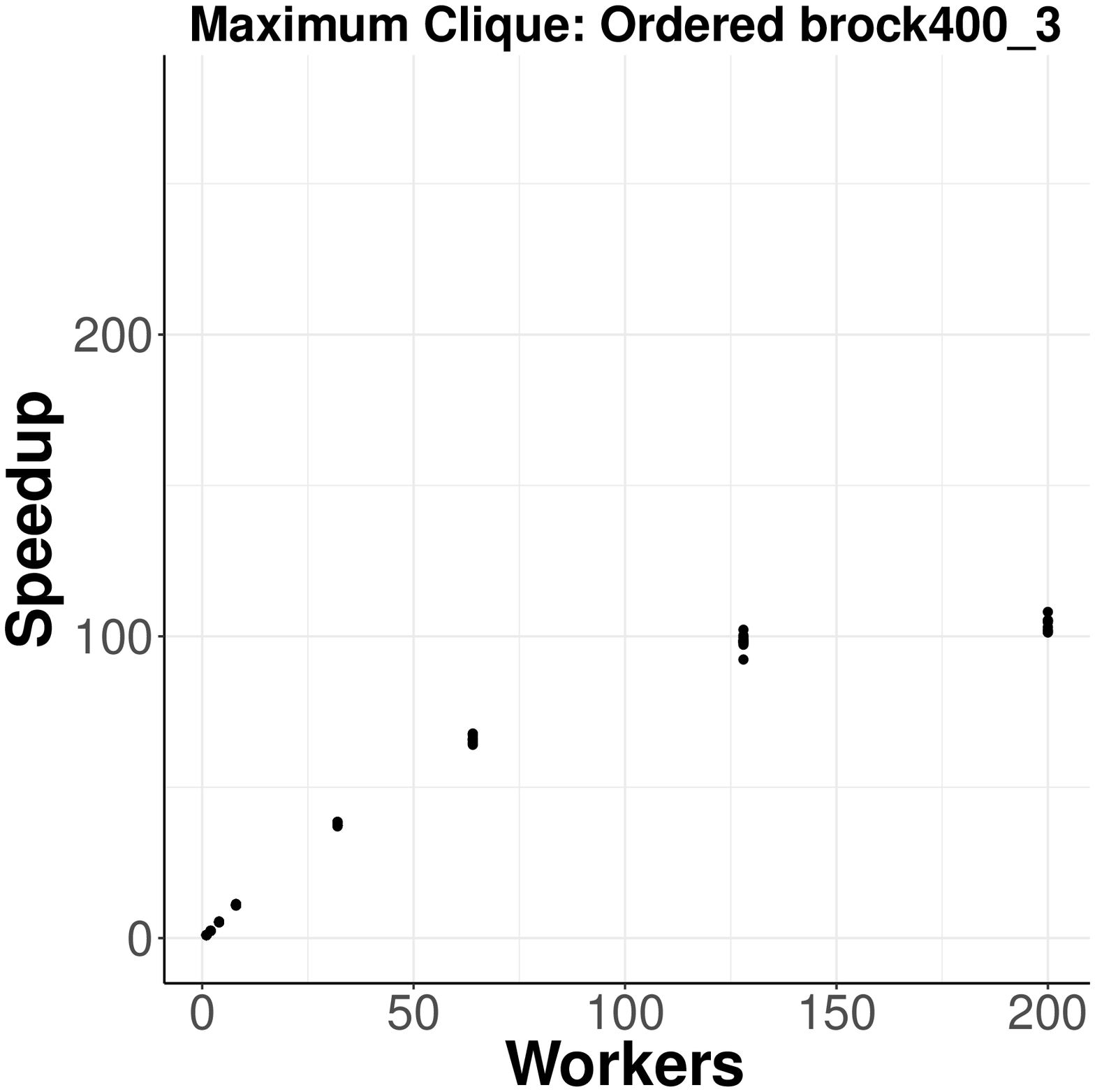}
  \end{subfigure}
  \begin{subfigure}[b]{0.3\textwidth}
    \includegraphics[width=\textwidth]{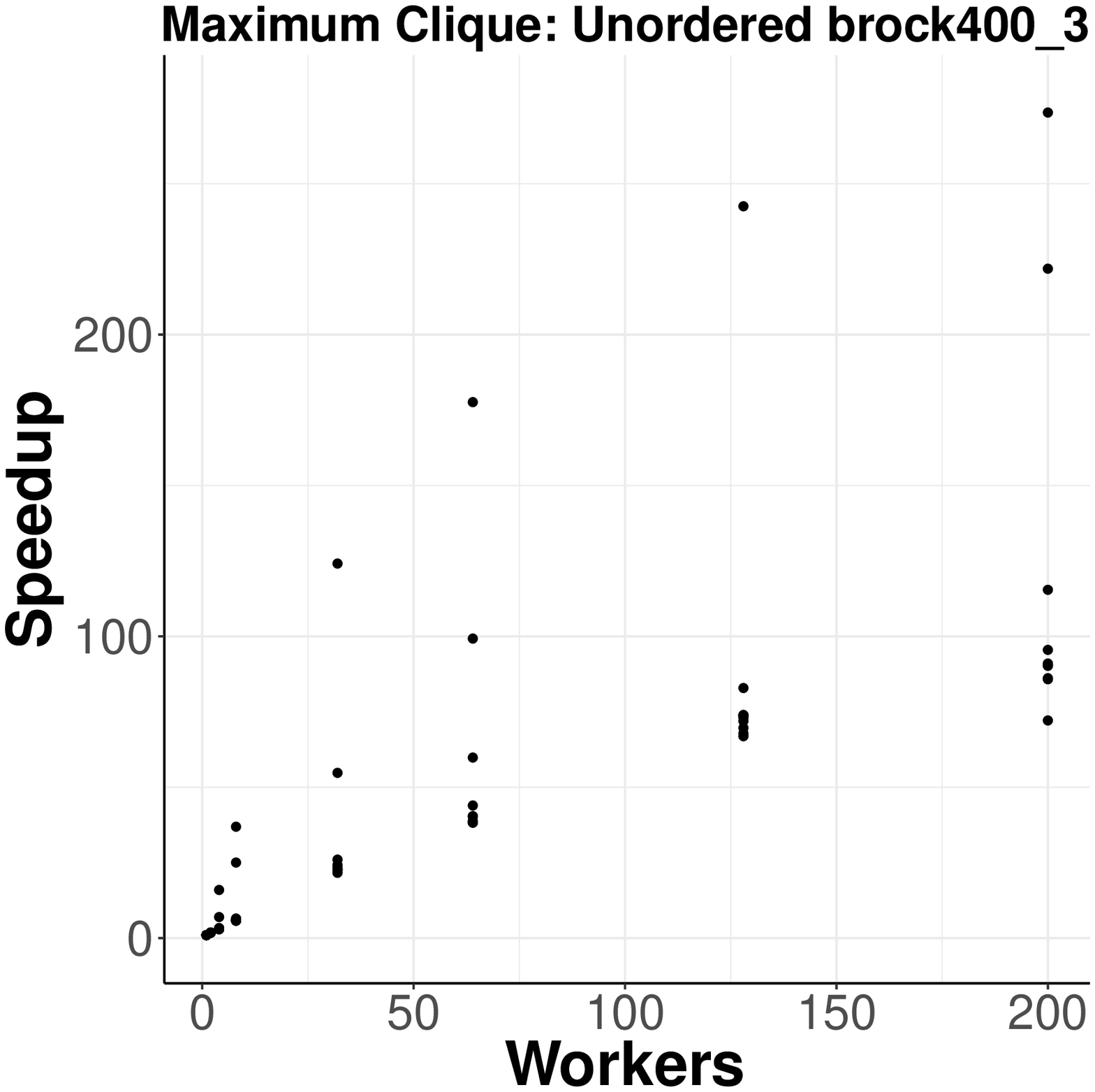}
  \end{subfigure}
  \caption{Unordered Skeleton Violates {\propB}: Sampled speedups for Maximum Clique}
  \label{fig:sampledScaling}
\end{figure}

While the mean speedups reported in \cref{fig:MC_Speedup} suggest that the
Unordered skeleton also preserves {\propB} they disguise the huge runtime
variance of the searches. \Cref{fig:sampledScaling} illustrates this by showing
each individual speedup sample from the brock400\_3 instance. The unpredictable
speedups for the Unordered skeleton are in stark contrast to the Ordered
skeleton. We attribute the high variance of the Unordered skeleton to the interaction between random
scheduling and search ordering.

\begin{figure}[h!]
  \centering
    \includegraphics[width=0.7\textwidth]{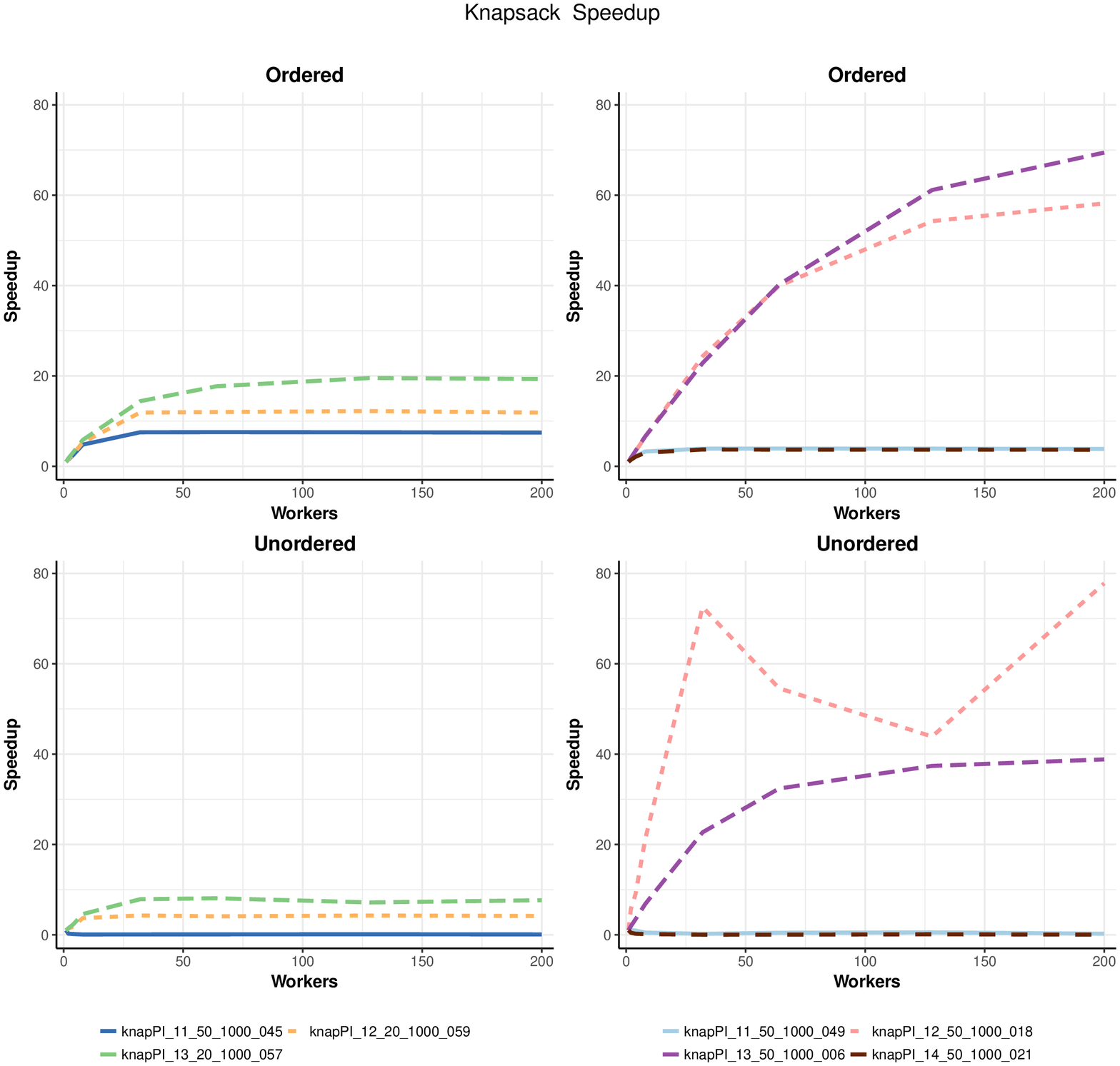}
  \vspace{-0.3cm}
  \caption{Knapsack Speedups: Ordered Skeleton Maintains {\propA} and
    {\propB} Properties}
  \label{fig:KP_Speedup}
\end{figure}

\begin{figure}[h!]
  \centering
  \includegraphics[width=0.7\textwidth]{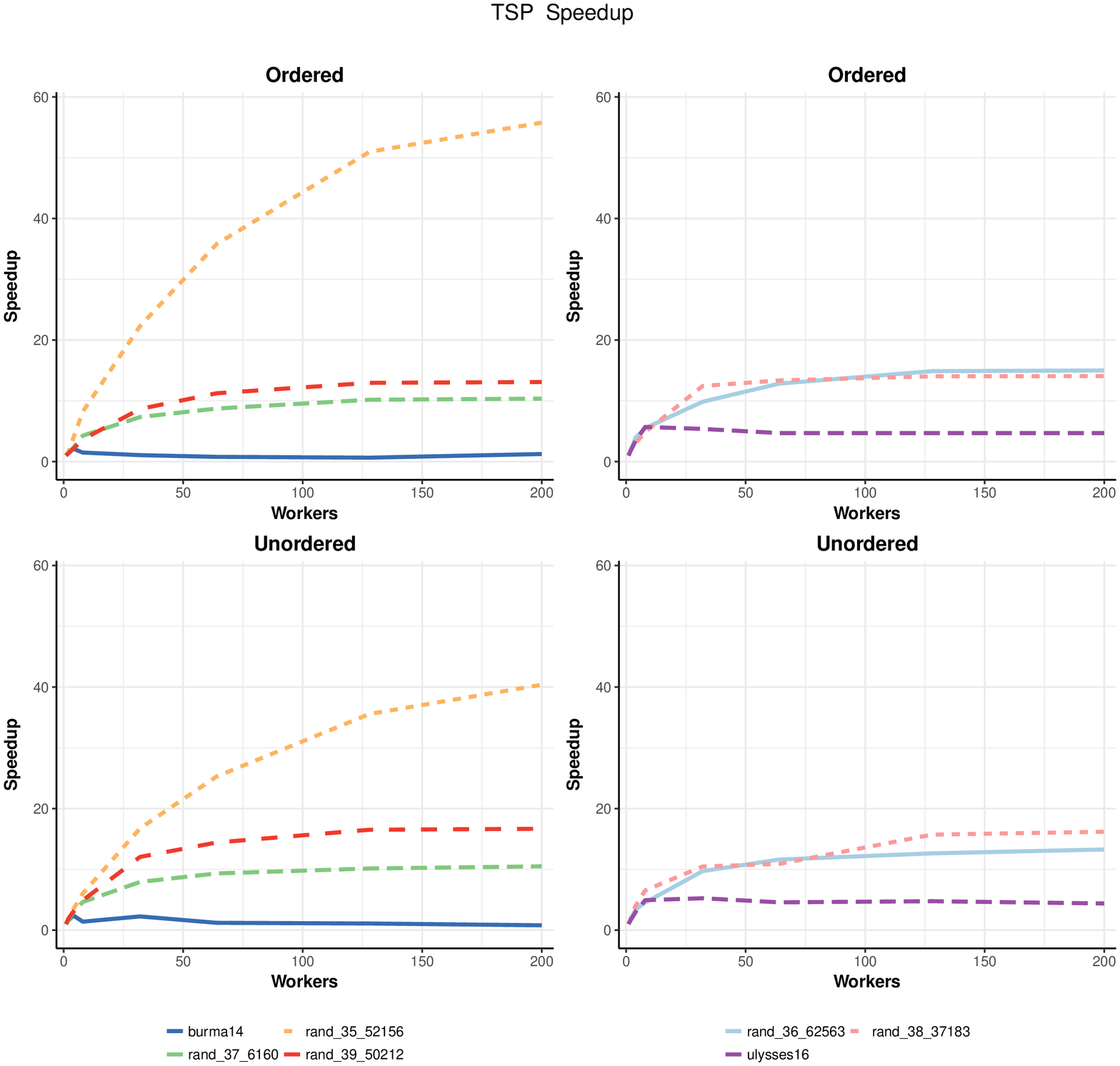}
  \vspace{-0.3cm}
  \caption{Travelling Salesperson Speedups: Ordered Skeleton Maintains {\propA} and {\propB} Properties}
\label{fig:TSP_Speedup}
\end{figure}

The speedup curves for the Knapsack and TSP applications are given in
\cref{fig:KP_Speedup} and \cref{fig:TSP_Speedup} respectively. Again
any instances with sequential runtimes greater than an hour are
excluded.

All travelling salesperson instances, for both skeletons, maintain
{\propA}\footnote{Short running times ($<$ 1s) for burma14 cause it to fail
  {\propA} in some cases, we put this down to runtime variance rather than
  ordering effects}. On Knapsack however, in contrast to the Ordered skeleton,
the Unordered skeleton deviates from {\propA} for five instances:
knapPI\_11\_50\_1000\_045 (2 -- 200 workers), knapPI\_11\_50\_1000\_049 (4 --
200 workers), knapPI\_14\_50\_1000\_021 (2 -- 200 workers),
knapPI\_15\_100\_1000\_059 (8 -- 200 workers) and knapPI\_15\_50\_1000\_072 (8
-- 200 workers) where the deviation is shown by a speedup of less than one
(\cref{fig:KP_Speedup}, bottom).

{\propB} is maintained by the Ordered skeleton in all Knapsack and Travelling
Salesperson cases except ulysses16 which shows a slowdown when moving from 32 to
64 workers. As with the brock400\_4 slowdown, the runtime at this scale is small
(10s), and deviations are likely caused by parallelism overheads rather than
search ordering effects.

While it is difficult to directly compare Ordered and Unordered executions due to search ordering effects, in some instances the Unordered skeleton is more efficient than the Ordered skeleton.
This is caused by a variety of factors including reduced warm-up time due to not requiring upfront work generation, distributed work stealing reducing the impact of a single node bottleneck and the potential for randomness to find a solution quicker than the fixed search order of the ordered skeleton.

\subsection{{\propC}}
\label{sec:repeatability}

The {\propC} property aims to ensure that multiple runs of the same
configuration have similar runtimes. To give a normalised variability measure we
use relative standard deviation (RSD)\footnote{Also known as coefficient of
  variation.}, i.e.\ the ratio of the standard deviation to the
mean~\cite{dictionaryOfStats}. To compare the variability of the benchmark
instances using the Ordered and Unordered skeletons we plot the RSD as a
cumulative distribution function (CDF) for each worker configuration. Here
the key metric is how quickly the curve reaches 1, i.e. the point that covers
all RSD values. A disadvantage of this type of plot is that it is not robust to
outliers. These plots contain all benchmarks including those where the
sequential run timed out but a parallel run was successful in less than an hour.
Benchmarks with mean runtime less than 5 seconds are removed as a high RSD is
expected.

\begin{figure}[h!]
  \centering
  \includegraphics[width=0.6\textwidth]{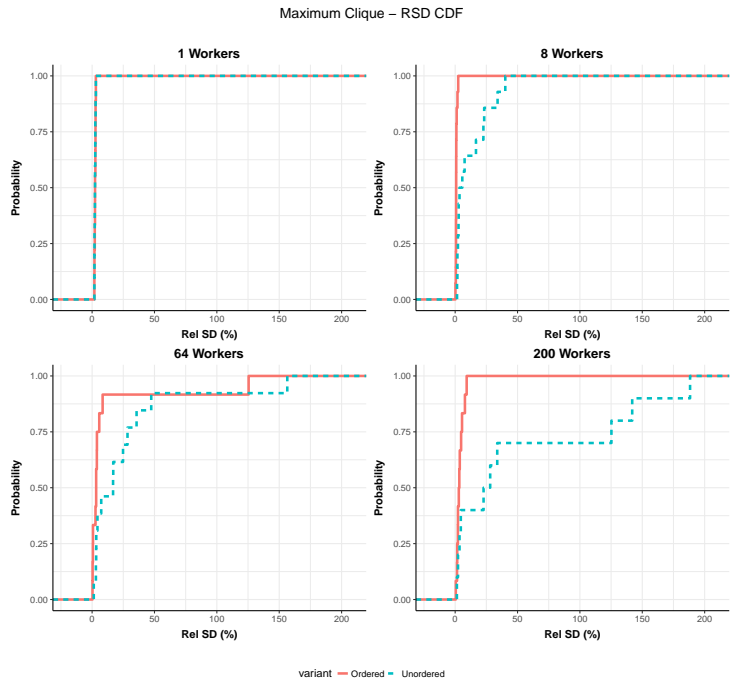}
  \vspace{-0.2cm}
  \caption{Maximum Clique Relative Variability (CDF of RSD): Ordered
    Skeleton Maintains {\propC} Property}
  \label{fig:maxclique-CDF}
\end{figure}

\Cref{fig:maxclique-CDF} shows the CDF plot for both skeletons for all maximum
clique benchmarks run with 1, 8, 64 and 200 workers. With a single worker the
maximum RSD of both skeletons is less than 3\% showing that they provide
repeatable results. This is expected as in the single worker case the
Unordered skeleton behaves like the Ordered skeleton, following a fixed
left-to-right search order.
With multiple workers the Ordered skeleton guarantees better
repeatability than the Unordered skeleton, with median RSDs given in
\cref{tab:medianRSDMC}. For the 64 worker case the long tails are caused by
outliers in the data and we see a low RSD maintained in almost 90\% of cases. The issues with identifying outliers are discussed in \cref{sec:dataOutlierExample}. The cause of
these outliers is unknown but, given the large discrepancy, is probably spurious
behaviour on the system rather than a manifestation of search order anomalies.

\begin{table}[h!]
  \centering
  \begin{tabular}{lrrrrrr}
    \toprule
    & \multicolumn{2}{c}{\textbf{Maximum Clique}} & \multicolumn{2}{c}{\textbf{Knapsack}} & \multicolumn{2}{c}{\textbf{TSP}}  \\
    \textbf{Workers} & \textbf{Ordered } & \textbf{Unordered } & \textbf{Ordered } & \textbf{Unordered } & \textbf{Ordered } & \textbf{Unordered } \\
    \midrule
    1   & 2.36 & 2.29   & 2.52 & 2.71   & 2.40 & 2.22 \\
    2   & 1.42 & 4.21   & 1.24 & 141.95 & 1.58 & 14.52 \\
    4   & 0.94 & 16.17  & 1.46 & 75.51  & 0.89 & 9.65 \\
    8   & 0.80 & 4.60   & 1.25 & 107.02 & 1.84 & 10.04 \\
    32  & 2.35 & 10.03  & 1.94 & 127.06 & 4.63 & 12.77 \\
    64  & 3.52 & 15.16  & 1.90 & 93.12  & 5.18 & 13.54 \\
    128 & 3.78 & 12.19  & 1.60 & 110.38 & 3.08 & 6.42 \\
    200 & 3.51 & 15.31  & 1.99 & 126.18 & 3.51 & 3.89 \\
    \bottomrule
  \end{tabular}
  \caption{Median Relative Standard Deviation (RSD) \% : Ordered Skeleton is more Repeatable}
  \label{tab:medianRSDMC}
\end{table}

\begin{figure}[h!]
  \centering
  \includegraphics[width=0.6\textwidth]{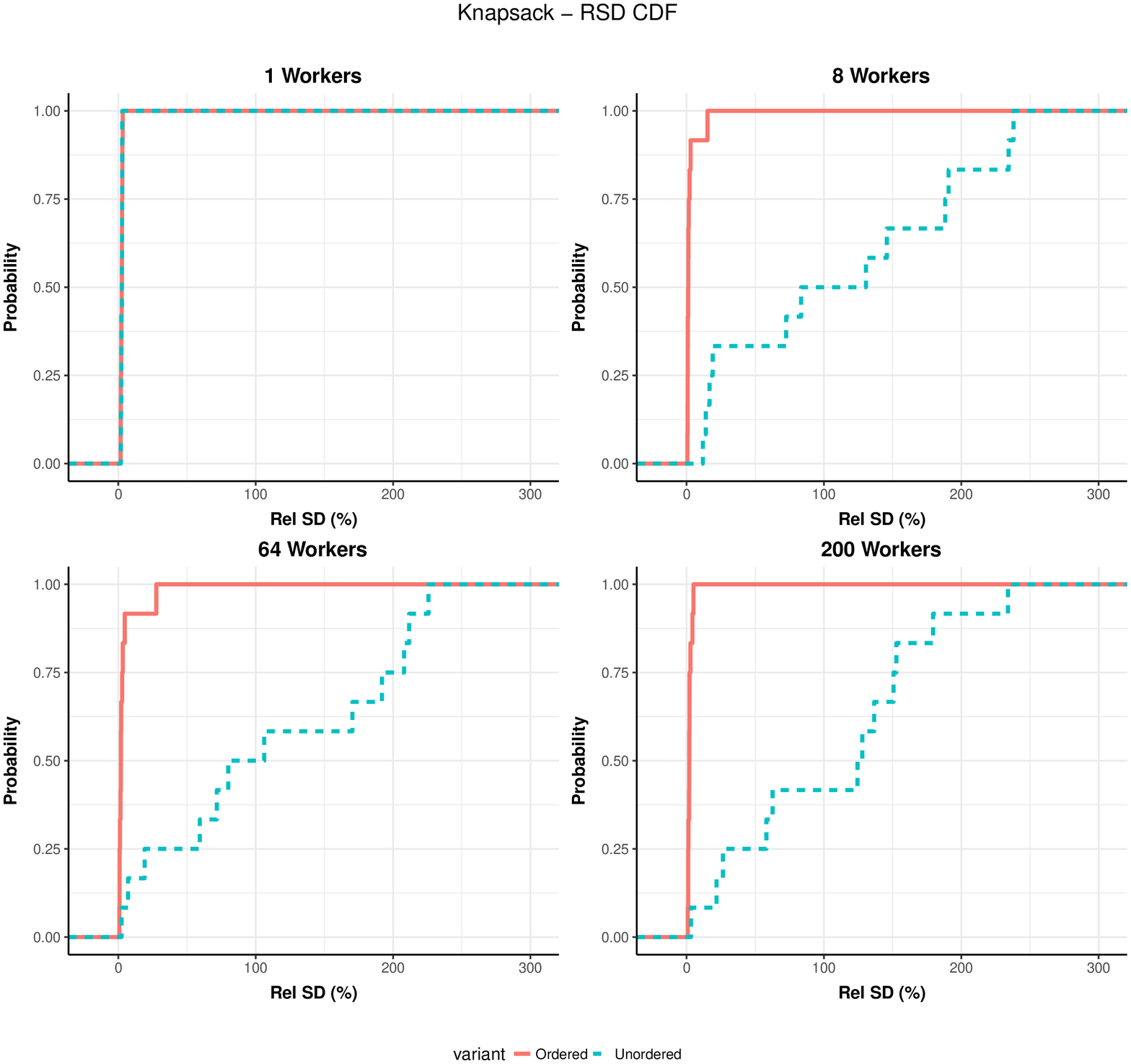}
  \vspace{-0.3cm}
  \caption{Knapsack Relative Variability (CDF of RSD): Ordered
    Skeleton Maintains {\propC} Property}
  \label{fig:knapsack-CDF}
\end{figure}

\begin{figure}[h!]
  \centering
  \includegraphics[width=0.6\textwidth]{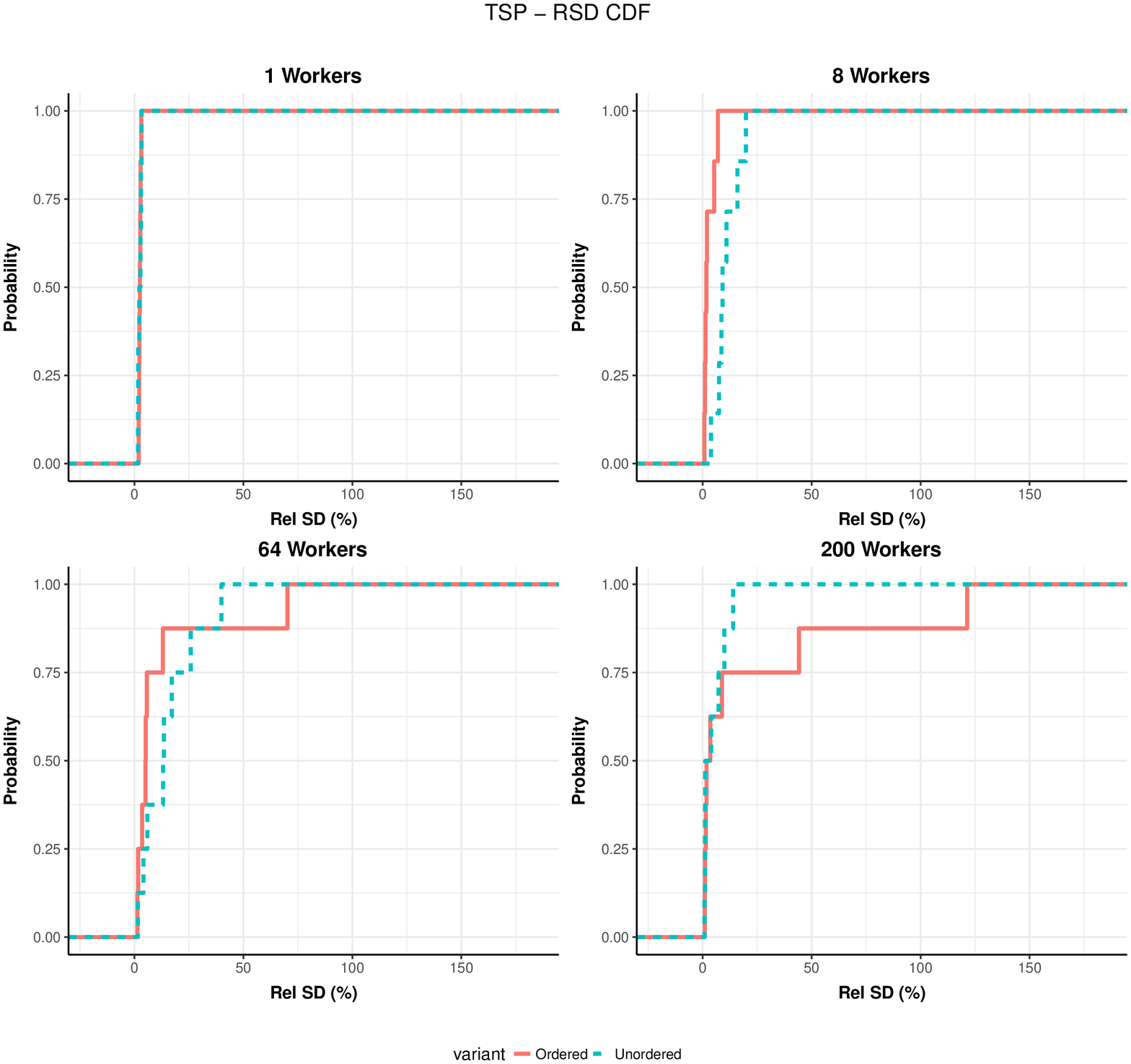}
  \vspace{-0.3cm}
\caption{TSP Relative Variability (CDF of RSD): Ordered Skeleton Maintains
  {\propC} Property. Long Tails are Caused by Data Outliers.}
  \label{fig:TSP-CDF}
\end{figure}

Figures \ref{fig:knapsack-CDF} and \ref{fig:TSP-CDF} show the CDF
plots for the knapsack and travelling salesperson benchmarks, and the
results are very similar to those for Maximum Clique. With a single
worker both Ordered and Unordered skeleton implementations deliver
highly repeatable results, i.e.\ a maximum RSD of less than 3\%. The
knapsack application has poor repeatability in the Unordered skeleton
cases; half of them suffering over 100\% RSD. As with Maximum Clique,
outlying data points make the Ordered skeleton appear to perform badly
on one or two of the TSP benchmarks in the the 64 and 200 worker
cases, as discussed in \cref{sec:dataOutlierExample}. Nonetheless the
Ordered skeleton maintains a low RSD.

\section{Conclusion}
\label{sec:conclusion}

Branch and bound searches are an important class of algorithms for
solving global optimisation and decision problems.
However, they are difficult to parallelise due to their sensitivity to
search order, which can cause highly variable and unpredictable
parallel performance.
We have illustrated these parallel search anomalies and propose that replicable
search implementations should avoid them by preserving three key
properties: {\propA}, {\propB} and {\propC}
(\cref{sec:motivatingExample}).
The paper develops a generic parallel branch and bound skeleton
and demonstrates that it meets these properties.

We defined a novel formal model for general parallel branch and bound
backtracking search problems (BBM) that is parametric in the search
order and models parallel reduction using small-step operational
semantics (\cref{sec:formalModel}).  The generality of the model was
shown by specifying three benchmarks: Maximum Clique, 0/1 Knapsack and
Travelling Salesperson.

We presented a Generic Branch and Bound (GBB) API as a set of higher
order functions (\cref{sec:genericBBSearch}). The GBB API conforms to
the BBM and its generality was shown by using it to implement the three
benchmarks. The Maximum Clique implementation is the first
\textit{distributed-memory} parallel implementation of San Segundo's bit
parallel Maximum Clique algorithm
(BBMC)~\cite{segundo.matia.ea_improvedBBMC:2011}.

We factored the elaborate parallel search behaviours as a pair of
sophisticated \emph{algorithmic skeletons} for distributed memory
architectures (\cref{sec:skeletons}).  While the \emph{Unordered
  skeleton} does not guarantee the parallel search properties the
\emph{Ordered skeleton} does. For example to guarantee the {\propA}
one thread is assigned to follow the sequential search order.

We have evaluated the parallel performance of the skeletons with 40
instances of the three benchmark searches
(\cref{tab:mcinstances,tab:tspInstances,tab:knapsackInstances}) on
a cluster with 17 hosts and up to 200 workers. The sequential performance of
the generic skeletons, implemented in Haskell, is between 1.9 and 6.2 times
slower than a  state of the art Maximum Clique solver implemented in C++.
The slowdown is primarily due to the relative execution speeds of Haskell
and C++, but also due to the generality of the skeletons and the lack
of hand optimisation (\cref{sec:seq-eval}).

We evaluated the properties using speedups relative to
the sequential runtime of the generic skeletons.
We find that the Ordered skeleton preserves the {\propA}
property for all benchmark instances with non-trivial runtimes. In contrast the
Unordered skeleton violates the property for 5 TSP instances
(\cref{sec:PropAPropB}). The Ordered skeleton preserves the {\propB} property
for all benchmark instances with non-trivial runtimes. In contrast the Unordered
skeleton violates the property for many instances of all three benchmarks
(\cref{sec:PropAPropB} and \cref{fig:sampledScaling}). The Ordered
skeleton delivers far more repeatable performance than the Unordered skeleton
with a median relative standard deviation of
1.78\% vs 5.56\%, 1.83\% vs 87.56\% and 1.96\% vs 8.61\% over all Maximum Clique,
Knapsack and TSP instances respectively (\cref{sec:repeatability}).

\subsubsection*{Acknowledgements}

This work is funded by UK EPSRC grants AJITPar (EP/L000687), CoDiMa (EP/M022641),
Glasgow DTA (EP/K503058), MaRIONet (EP/P006434), Rathlin (EP/K009931)
and Border Patrol (EP/N028201/1).
We also thank the anonymous reviewers for their feedback.

\section*{References}
\bibliographystyle{elsarticle-num}
\bibliography{paper}

\appendix

\section{Repeatability and Data Outliers}
\label{sec:dataOutlierExample}

\begin{table}[h]
  \centering
\begin{tabular}{rr}
  \toprule
  \textbf{Unordered} & \textbf{Ordered} \\
  \midrule
  120.4 & 191.8 \\
  159.1 & 192.0 \\
  183.0 & 194.1 \\
  183.6 & 197.1 \\
  201.7 & 197.9 \\
  216.6 & 198.2 \\
  220.4 & 202.7 \\
  246.7 & 206.9 \\
  303.4 & 207.3 \\
  \textbf{619.5} & \textbf{444.0} \\
  \bottomrule
\end{tabular}
  \caption{Runtime Outliers in a 32 Worker TSP Instance}
  \label{tab:outliers}
\end{table}

\cref{tab:outliers} illustrates some of the issues with outliers in
the runtime measurements. The dataset is from the
rand\_34\_22137 TSP instance with 32 workers. The Ordered skeleton
runtimes have a potential outlier (in bold) with a runtime 237s
greater than any of the other runtimes. The other 9
runtimes have a range of just 15.5s (i.e.\ 207.3-191.8). In this case
it is almost certain that the outlier is not due to some search order
effect, but rather to some external system factor, e.g.\ network
contention or some daemon process running. We have not been able to reproduce
this effect in additional experiment runs.

The Unordered skeleton runtimes also appear to have an outlier (in
bold) with a runtime 316.1s greater than any of the other runtimes. It
is, however, harder to be certain that this is an external factor as
the variability of the other 9 measurements is far higher, i.e. 183s
(303.4-120.4).

As a result of the difficulties of identifying them we have not
attempted to eliminate any outliers, even where there is a strong
case. That is, the cumulative distribution function (CDF) plots in
\cref{fig:maxclique-CDF,fig:knapsack-CDF,fig:TSP-CDF} show all
measurements recorded. In addition we use median relative standard
deviation (RSD) to eliminate the effects of any outliers when
comparing the repeatability of the Ordered and Unordered skeletons
(\cref{tab:medianRSDMC}).

\section{0/1 Knapsack -- BBM Formalisation}
\label{sec:formalKnapsack}

The input parameters for the 0/1 knapsack problem consists of a fixed capacity \(C\) and a
set of \(m\) items \(I = \{i_1 \dots i_m\}\). For each item \(i_1 \in I\) we
define \(p : I \to \mathbb{N}\) and \(w : I \to \mathbb{N}\) to be projection
functions giving the profit and weight of an item respectively.
If  the input is considered as a vector of items, then the  problem
is equivalent to creating an output vector \(X\) where \(x_j \in \{0,1\}\),
encoding the inclusion or exclusion of each item \(i_j\) in the input vector:

\begin{equation}
  \textbf{max} \sum_{j=1}^{m} p(i_j)x_j \quad \textbf{where}\; \sum_{j=1}^{m} w(i_j)x_j \leq C
\end{equation}

The BBM model  uses finite words to represent tree branches. We can trivially extend
the profit and weight functions, \(p\) and \(w\), to those operating on words
(\(p : I^* \to \mathbb{N}\) and \(w : I^{*} \to \mathbb{N}\))
by taking the sum across each component. An unordered tree generator, \(g : I^* \to 2^I\), takes the current set of items and  chooses any item not previously selected:

\begin{equation}
g(i_1 \dots i_m) = \{ i_j \in I \setminus \{i_1 \dots i_m\} ~|~ w(i_1 \dots i_m) + w(i_j) \leq C \}
\end{equation}

One ordering heuristic is \emph{profit density}, and an ordered generator
simply orders the results of the unordered generator by ascending
profit density.

\begin{equation}
\frac{p(i_1)}{w(i_1)} \geq \frac{p(i_2)}{w(i_2)} \dots \geq \frac{p(i_n)}{w(i_n)}
\end{equation}

As the aim is to \emph{maximise} total profit the objective function,
\(f : I^* \to \mathbb{N}\) is simply \emph{p}, the profit function
over words. The ordering on objective functions \(\sqsubseteq\) is
given by the natural ordering \(\leq\).

Finally, the pruning predicate, \(p : \mathbb{N} \times I^* \to
\{0,1\}\), is defined as follows, where \emph{fractionalKnapsack}
greedily solves the residual knapsack problem using \emph{continuous
  relaxation} to obtain an optimal solution and hence a bound on the
maximal profit.

\begin{equation}
  p(bnd,(i_1 \dots i_m)) =
  \begin{cases}
    1 & \textbf{if}\; p(i_1 \dots i_m) + fractionalKnapsack(I \setminus \{i_1 \cdots i_m\}, C - w(i_1 \dots i_m)) \leq bnd \\
    0 & otherwise
  \end{cases}
\end{equation}

\section{0/1 Knapsack -- GBB and Skeleton Representation}
\label{sec:GBBknapsack}

\begin{haskellcode}{Knapsack in the GBB API}{lst:GBBKnapsack}
type Space           = (Array Int Int, Array Int Int)
type Profit          = Int
type Wight           = Int
type Candidates      = [Item]
data PartialSolution = Solution Capacity [Item] Profit Weight

type KPNode          = (PartialSolution, Profit, Candidates)

orderedGenerator :: Space -> KPNode -> [KPNode]
orderedGenerator items (Solution cap is currentProfit currentWeight, bnd, remaining) =
  map createNode (filter (\item -> currentWeight + (itemWeight items item) <= cap) remaining)
  where
    createNode :: Item -> KPNode
    createNode i = let
      newSol   = Solution
                   cap
                   (i:is)
                   (currentProfit + itemProfit items i)
                   (currentWeight + itemWeight items i)
      newBnd   = currentProfit + (itemProfit items i)
      newCands = delete i remaining
      in (newSol, newBnd, newCands)

pruningHeuristic :: Space -> KPNode -> Int
pruningHeuristic items (Solution cap (i:is) solP solW, _, _) =
  round $ fractionalKnapsack items solP solW (i + 1)

-- Defined elsewhere. Solve knapsack allowing for fractional values
fractionalKnapsack :: Space -> Profit -> Weight -> Int -> Double

-- Calling a skeleton implementation
Unordered.search
  spawnDepth
  (KPNode (Solution cap items 0 0, 0, items))
  orderedGenerator
  pruningHeuristic
\end{haskellcode}

A Knapsack implementation using the GBB API and the Unordered skeleton is shown
in \cref{lst:GBBKnapsack}. The implementation comes directly from BBM,
generating only candidate items which do not exceed the capacity constraint and
using continuous relaxation to compute an upper bound. The
\texttt{Unordered.search} function on line 32 invokes the Unordered skeleton
implementation with an empty root node. The Ordered skeleton invocation is very
similar. A higher performance version supporting distributed execution is used
for evaluating the skeletons in \cref{sec:eval-knapsack}.

\section{Travelling Salesperson -- BBM Formalisation}
\label{sec:formalTSP}

The TSP input consists of a set \(C\) of \(n\) cities and a metric on \(C\), given
by a symmetric non-negative distance function \(d : C \times C \to \mathbb{R}\).

Tours are modelled as words over \(C\) where
the word \(t = c_1 c_2 \dots c_k \in C^*\) represents a \emph{(partial) tour}
starting at \(c_1\) and ending at \(c_k\) if all cities \(c_i\) are pairwise
distinct.
The tour \(t\) is \emph{complete} if \(k = n\),
that is, if every city in \(C\) is visited exactly once.

We generalize the distance function \(d\) to words in \(C^*\) in the obvious
way:
\begin{align*}
  d(\epsilon)              & = 0 \\
  d(c_1)                   & = 0 \\
  d(c_1 \dots c_k c_{k+1}) & = d(c_1 \dots c_k) + d(c_k, c_{k+1})
\end{align*}

The (unordered) tree generator function, \(g: C^* \to 2^C\), extends a partial
tour with each city that hasn't been visited yet, enumerating all possible
tours. Due to symmetries --- rotations to change the start, reflections to
change the direction --- each tour is enumerated \(2n\) times. This symmetry can
be broken by fixing the starting city.

\begin{align*}
g(c_1 \dots c_k) & = C \setminus \{c_1,\dots,c_k\}
\end{align*}

The objective function, \(f: C^* \to \mathbb{R}\), maps complete tours to
the total distance travelled (including the distance from the last city
back to the starting city) and incomplete tours to infinity~\footnote{Formally, the co-domain of \(f\) should be
  \(\mathbb{R} \cup \{\infty\}\); we ignore this detail.
  In practice, \(\infty\) can be replaced by any real number larger
  than the total distance of the longest possible tour.}:

\begin{align*}
  f(c_1 \dots c_k) & =
  \begin{cases}
    d(c_1 \dots c_k c_1) & \text{if \(c_1 \dots c_k\) is a complete tour} \\
    \infty               & \text{otherwise}
  \end{cases}
\end{align*}
We aim to \emph{minimise} the objective function with respect
to the standard order $\leq$ on the reals; in BBM this corresponds to
maximising $f$ with regard to the dual order $\geq$ on $\mathbb{R}$  where $\infty$ is the minimal element \wrt $\geq$.

Finally, the pruning predicate, \(p : \mathbb{R} \times C^* \to \{0,1\}\),
prunes a partial tour if the distance travelled
along the tour plus the weight of a minimum spanning tree covering
the remaining cities exceeds the distance of the current shortest tour:
\begin{align*}
  p(\mathit{minDist}, c_1 \dots c_k) =
  \begin{cases}
    1 & \text{if}~d(c_1 \dots c_k) +
        \mathit{weightMST}(C \setminus \{c_2,\dots,c_{k-1}\}) \geq
        \mathit{minDist} \\
    0 & \text{otherwise}
  \end{cases}
\end{align*}
Here, \(\mathit{weightMST}(C \setminus \{c_2,\dots,c_{k-1}\})\)
is the weight of a minimum spanning tree covering the not-yet visited cities
as well as the starting city \(c_1\) and the most recently visited city \(c_k\).
The weight of this MST is a lower bound of the distance covered in  the
shortest partial tour from \(c_k\) through the not-yet visited
cities in \(C \setminus \{c_1,\dots,c_k\}\) and back to the start \(c_1\).

\section{Travelling Salesperson -- GBB and Skeleton Representation}
\label{sec:GBBTSP}

\begin{haskellcode}{TSP in the GBB API}{lst:GBBTSP}
type City       = Int
type Path       = [City]
type Candidates = IntSet
type Solution   = (Path, Int)
type TSPNode    = (Solution, Int, Candidates)

orderedGenerator :: DistanceMatrix -> TSPNode -> [TSPNode]
orderedGenerator distances ((path, pathLen), bnd, remainngCities) =
  map constructNode remainngCities
  where
    constructNode :: City -> TSPNode
    constructNode city =
      let newPath  = path ++ city
          newDist  = pathLen + distanceBetween (last path) city
          newRemainngCities   = delete city remainngCities
      in
        if not (null newRemainngCities) then
          ((newPath, newDist), bnd, newRemainngCities)
        else
         -- Only update the bound when we have a complete path
         let newPath'  = newPath ++ first path
             newDist'  = newDist + distanceBetween (last newPath) (first path)
         in ((newPath', newDist'), newDist', [])

pruningHeuristic :: DistanceMatrix -> TSPNode -> Int
pruningHeuristic dists ((path, pathLen), bnd, remainngCities) =
  pathLen + weightMST dists (last path) (insert (first path) remainngCities)

-- Defined elsewhere. Compute the minimum spanning tree cost via Prim's algorithm
weightMST :: DistanceMatrix -> City -> [City] -> Int

-- Calling a skeleton implementation
Unordered.search
  spawnDepth
  (TSPNode (([1],0), greedyNearestNeighbour cities, delete 1 cities))
  orderedGenerator
  pruningHeuristic
\end{haskellcode}

A TSP implementation using the GBB API and Unordered skeleton is shown in
\cref{lst:GBBTSP}. Unlike in the Maximum Clique and Knapsack benchmarks, bound
updates only happen when there are no cities left to choose which requires
additional logic in the \texttt{orderedGenerator} function. When calling the
skeleton on line 33, the root node sets an initial solution with the initial
city selected single city selected. The bounds are initially set to the result
of a greedy nearest neighbour algorithm to improve early pruning.

\end{document}